\documentclass[10pt,twocolumn,letterpaper,x11names]{article}

\usepackage{cvpr}              

\usepackage{booktabs}
\usepackage{multirow}
\usepackage{multicol}
\usepackage{colortbl}
\usepackage{gensymb}

\usepackage{float}

\definecolor{cvprblue}{rgb}{0.21,0.49,0.74}
\usepackage[pagebackref,breaklinks,colorlinks,allcolors=cvprblue]{hyperref}

\newcommand{\name}{Vid2Sim\xspace}

\def\paperID{2959} 
\def\confName{CVPR}
\def\confYear{2025}

\title{\name: Generalizable, Video-based Reconstruction of Appearance, Geometry and Physics for Mesh-free Simulation}

\author{Chuhao Chen$^{1}$ \quad Zhiyang Dou$^{1,2}$ \quad Chen Wang$^{1}$ \quad Yiming Huang$^{1}$ \\
\quad Anjun Chen$^{1,3}$ \quad Qiao Feng$^{1}$ \quad Jiatao Gu$^{1}$ \quad Lingjie Liu$^{1}$ \\ 
\vspace{2mm}
$^1$University of Pennsylvania \quad 
$^2$The University of Hong Kong \quad $^3$Zhejiang University \\
\vspace{2mm}
{\tt\normalsize \{chuhaoc,zydou,chenw30,ymhuang9,chen3110,fengqiao,jgu32,lingjie.liu\}@seas.upenn.edu} \\ 
\url{https://czzzzh.github.io/Vid2Sim} 
\vspace{2mm}
}

\begin{document}
\twocolumn[{
 \renewcommand\twocolumn[1][]{#1}
\maketitle
\begin{center}
    \vspace{-8mm}
    \captionsetup{type=figure}
     \includegraphics[width=\linewidth]{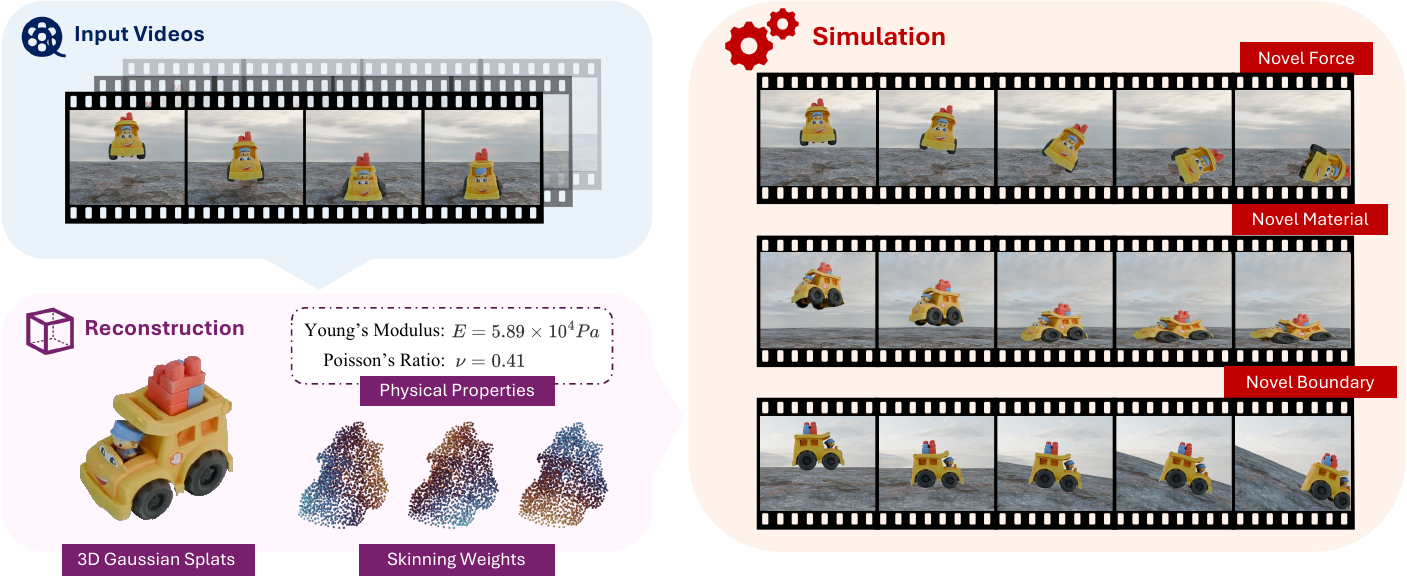}
    \captionof{figure}{\name achieves high-quality reconstruction of appearance, geometry, and physics from multi-view videos effectively. The reconstruction results are simulation-ready, enabling high-fidelity and visually appealing animations via mesh-free simulation. Here, we present our method's reconstruction and simulation results in the GSO~\cite{downs2022google} dataset.}
    \label{fig:teaser}
\end{center}}
]
\begin{abstract}
Faithfully reconstructing textured shapes and physical properties from videos presents an intriguing yet challenging problem. Significant efforts have been dedicated to advancing such a system identification problem in this area. Previous methods often rely on heavy optimization pipelines with a differentiable simulator and renderer to estimate physical parameters. However, these approaches frequently necessitate extensive hyperparameter tuning for each scene and involve a costly optimization process, which limits both their practicality and generalizability. In this work, we propose a novel framework, Vid2Sim, a generalizable video-based approach for recovering geometry and physical properties through a mesh-free reduced simulation based on Linear Blend Skinning (LBS), offering high computational efficiency and versatile representation capability. Specifically, Vid2Sim first reconstructs the observed configuration of the physical system from video using a feed-forward neural network trained to capture physical world knowledge. A lightweight optimization pipeline then refines the estimated appearance, geometry, and physical properties to closely align with video observations within just a few minutes. Additionally, after the reconstruction, Vid2Sim enables high-quality, mesh-free simulation with high efficiency. Extensive experiments demonstrate that our method achieves superior accuracy and efficiency in reconstructing geometry and physical properties from video data.
\end{abstract}    
\vspace{-5mm}
\section{Introduction}
\label{sec:introduction}
Understanding and reconstructing appearance, geometry, and physical properties from observations with high fidelity, a.k.a. system identification, is a fundamental yet challenging task in computer vision. Traditional methods~\cite{qiao2021differentiable, rojas2021differentiable, du2021diffpd, geilinger2020add, heiden2021disect, jatavallabhula2021gradsim, ma2022risp} often rely on known shape information of given objects, which limits their practicality for broader applications. Recent advancements~\cite{li2023pac,kaneko2024improving,zhong2025reconstruction,cai2024gaussian} leverage neural representations, such as NeRF~\cite{mildenhall2021nerf} and Gaussian Splatting~\cite{kerbl20233d} along with differentiable simulators~\cite{jiang2016material} to create a unified framework that jointly learns 3D geometry, appearance, and physical parameters. That being said, none of the previous efforts have achieved accurate, generalizable, and efficient reconstruction of appearance, geometry, and physical properties from the input video, as they suffer from two main limitations. First, most existing methods~\cite{li2023pac,kaneko2024improving,zhong2025reconstruction,cai2024gaussian} employ heavy per-scene optimization to identify physical parameters, making the understanding of various scenes computationally expensive. Second, these approaches struggle to accurately model complex, physics-driven deformations, as they typically use Material Point Methods (MPM) \cite{jiang2016material} for simulation. This method is limited by its grid-based representation and its typical dependence on symplectic time integration, which constrains expressiveness. Although alternative approaches, such as Spring-Gaus \cite{zhong2025reconstruction}, employ more efficient mass-spring models, they are limited to modeling elastic dynamics.

In this paper, we propose a novel framework, named \name, for the high-fidelity reconstruction of textured shapes and the estimation of physical properties directly from videos. We first train a feed-forward neural network that integrates general physical knowledge, utilizing a pre-trained video vision transformer~\cite{tong2022videomae} to infer a range of physical attributes from the input video sequences. This component is coupled with an advanced 3D reconstruction pipeline~\cite{tang2025lgm} that predicts both object geometry and appearance, encoded with 3D Gaussians to facilitate instant system identification. In contrast to prior methods, \name incorporates an efficient simulation pipeline leveraging an implicit Euler solver as inspired by~\cite{modi2024simplicits}. This simulation approach is mesh-free and uses Linear Blend Skinning (LBS) to enable reduced-order, computationally efficient simulations that are highly adaptable to complex deformations and fully end-to-end trainable. Then, we perform a lightweight optimization with a novel Neural Jacobian module to efficiently refine estimates of appearance, geometry, and physical properties, aligning the reconstructed outputs precisely with observed video data. This post-prediction optimization completes in only a few minutes. Upon reconstruction, the system enables high-quality, mesh-free simulations via the implicit Euler solver, supporting accurate dynamic behavior modeling.

We conduct extensive experiments to evaluate our method where \name demonstrates remarkable accuracy and efficiency in recovering geometry, appearance, and physical properties from videos compared to existing methods. In summary, our contributions are three-fold:
\begin{itemize}
\item  We propose \name, a novel framework for generalizable, video-based reconstruction of appearance, geometry, and physical properties for mesh-free, reduced-order simulation.
\item We introduce a generalizable feed-forward model with physical world knowledge to estimate the dynamics, followed by an efficient optimization step with Neural Jacobian to improve the reconstruction results further.
\item \name demonstrates remarkable effectiveness and efficiency, achieving state-of-the-art performance in accuracy and speed compared to existing methods.
\end{itemize}

\section{Related Work}
\label{sec:related_work}

\subsection{Physics-aware Dynamic 3D reconstruction} 
Dynamic 3D reconstruction is one of the critical tasks in computer vision and graphics. Recent advances in 3D representations like NeRF \cite{mildenhall2021nerf} and 3D Gaussian Splatting \cite{kerbl20233d} as well as template-based models~\cite{loper2023smpl, romero2022embodied, li2017learning} make it flexible to reconstruct complex 3D scenes from visual data. These methods are recently extended to a dynamic 3D reconstruction~\cite{pumarola2021d, yang2023real, wu20244d} from either monocular videos \cite{xian2021space, gao2021dynamic, tretschk2021non, qiao2022neuphysics, yang2024deformable, wu2024dice, shimada2023decaf} or multi-view videos \cite{park2021nerfies, park2021hypernerf, luiten2023dynamic}. With the introduction of physics-informed learning \cite{cuomo2022scientific, chiu2022can}, approaches that incorporate physical priors to enhance the understanding and reconstruction of dynamic scenes have gained popularity. For instance, PAC-NeRF~\cite{li2023pac} first jointly reconstructed the dynamic scene and a simulatable model using the differentiable Material Point Method~\cite{hu2018moving, hu2019difftaichi}, and it was subsequently improved regarding the quality \cite{cai2024gaussian, kaneko2024improving} and adaptability \cite{zhong2025reconstruction}. While these methods achieve physically complete reconstruction, none of them are generalizable. In contrast to all existing methods, we first propose a generalizable pipeline that achieves simulation-ready geometry and physical property recovery in a feed-forward manner, which is inspired by the recent achievements in large 3D reconstruction model \cite{hong2023lrm, tang2025lgm, zhang2025gs} and 4D reconstruction model \cite{ren2024l4gm}. A highly efficient optimization step is conducted to further enhance the reconstruction quality. 

\subsection{Vision-based Physical Simulation}
\paragraph{Mesh-free Physical Simulation}
Traditional physical elasticity simulation, such as the finite element method (FEM) \cite{cutler2002procedural}, often requires a mesh or tetrahedral representation. This complicates the simulation of scenes reconstructed from visual data, often represented by NeRF or 3D Gaussians, as obtaining high-quality meshes from these models for simulation can be a non-trivial task. Mesh-free models have then been a popular alternative for vision-based physical simulation such as the material point method (MPM) \cite{jiang2016material, hu2018moving} and smoothed-particle hydrodynamics (SPH) \cite{desbrun1996smoothed, peer2018implicit, kugelstadt2021fast}. However, neither is a purely point-based method since SPH needs to update connectivity among neighborhoods and MPM requires maintaining a background grid. More importantly, these approaches bring significant computational burden. The very recent work Simplicits \cite{modi2024simplicits} thus proposed a mesh-free, geometry-agnostic, and reduced-order elastic simulation method, which offers another feasibility to do a vision-based physical simulation in an efficient and flexible way. Inspired by Simplicits \cite{modi2024simplicits}, we develop a feed-forward model that efficiently delivers a generalizable initial estimate, coupled with a differentiable, reduced-order simulator that employs Linear Blend Skinning for rapid and accurate optimization of appearance, geometry, and physical properties.

\paragraph{Physical reconstruction and simulation from visual data}
Apart from physics-aware dynamic 3D reconstruction, there are a lot of other applications in vision-based physical simulation with the help of mesh-free simulation methods. Works such as PhysGaussian \cite{xie2024physgaussian} integrate mesh-free simulators with NeRF \cite{feng2024pie} or 3D Gaussians \cite{jiang2024vr, lu2025manigaussian}, making it possible to interact with these representations. Some other works \cite{zhangphysics, liu2024physics3d, liu2025physgen, feng2024elastogen} combine the simulation model with the video generation model \cite{bar2024lumiere, blattmann2023stable, rombach2022high, brooks2024video} to learn physical properties and generate dynamics.
As of yet, all previous methods are limited by their reconstruction accuracy, generalization capability, and runtime cost.

\section{Preliminary}
We begin by introducing (1) mesh-free simulation~\cite{modi2024simplicits}, which operates without mesh or grid representation using a reduced-order simulator; and (2) 3D Gaussian Splatting~\cite{kerbl20233d} for modeling both geometry and appearance.

\label{sec:preliminary}
\paragraph{Mesh-Free, Reduced-Order Simulation}
Given a set of points $\{\mathbf{X}_i \in \mathbb{R}^3 ~|~ i=1, 2, ..., n\}$ at the rest position, following~\cite{modi2024simplicits}, we simulate the dynamics of the points with a set of handles (full affine transformations) $\{\mathbf{Z}_j \in \mathbb{R}^{3 \times 4} ~|~ j=1, 2, ..., m\}$ (or $\mathbf{z}_j \in  \mathbb{R}^{12}$ in an equivalent vector form) with a reduced $m \ll n$. The deformation of the point $\mathbf{X}_i$ is then defined as
\begin{equation}
    \mathbf{x}_i=\mathbf{\phi}_{i}(\mathbf{X}_i, \mathbf{Z})=\mathbf{X}_i+\sum_{j=1}^{m} W_{\theta;j}(\mathbf{X}_i)\mathbf{Z}_{j} [\mathbf{X}_i, 1]^\top,
    \label{eq:lbs}
\end{equation}
where $\mathbf{x}_i$ represents the deformed position, and $W_{\theta;j}(\mathbf{X}_i)$ is a scalar weight for Linear Blending Skinning~(LBS), predicted by a small Multilayer Perception (MLP) that models the transformation of each point based on the combined influence of the handles.

The handles $\mathbf{z}_i$ are initialized to zero to make sure the points are at the rest position at $t=0$. Then, at each discrete time step, the handles vary according to the implicit time integration with the following incremental potential equation containing an inertia term and a potential energy term:
\begin{equation}
\mathbf{z}_{t+1}=\mathop{\rm{argmin}}\limits_{\mathbf{z}} \frac{1}{2} \|\mathbf{z} - \mathbf{\tilde z}_t \|_{\mathbf{M}}+\Delta t^2 E_{\rm{potential}}(\mathbf{z}_t)
\end{equation}
where $\Delta t$ is the simulation time step, $\mathbf{\tilde z}_t=\mathbf{z}_t+\Delta t \mathbf{\dot{z}}_t$ is the first order prediction of $\mathbf{z}_t$ and $E_{\rm{potential}}(\mathbf{z}_t)$ is the potential energy from both internal and external forces. Following~\cite{modi2024simplicits}, when evolving $\mathbf{z}_t$ at each timestep, we usually sample a small set of key control points $\{\mathbf{X}^c_i \in \mathbb{R}^3 ~|~ i=1, 2, ..., k\},k \ll n$, which is also called \textit{cubature points}, to save the computational time and memory.

\paragraph{3D Gaussian Splatting}
3D Gaussian Splatting \cite{kerbl20233d} represents 3D scenes as Gaussian primitives. Each primitive is defined by the Gaussian function:
\begin{equation}
\mathcal{G}(\mathbf{x}) = e^{-\frac{1}{2}(\mathbf{x} - \mathbf{p})^\top \mathbf{\Sigma}^{-1} (\mathbf{x} - \mathbf{p})}
\end{equation}
where $\mathbf{p}$ is the center and $\mathbf{\Sigma} = \mathbf{R}\mathbf{S}\mathbf{S}^\top \mathbf{R}^\top$ is the covariance matrix, factorized into rotation matrix $\mathbf{R}$ and scaling matrix $\mathbf{S}$. For rendering, learnable parameters $\mathbf{p}$ and $\mathbf{\Sigma}$ are projected into camera coordinates as $\mathbf{p}' = \mathbf{K} \mathbf{W} [\mathbf{p}, 1]^\top, \mathbf{\Sigma}' = \mathbf{J} \mathbf{W} \mathbf{\Sigma} \mathbf{W}^\top \mathbf{J}^\top$, where $\mathbf{K}$ is the camera's intrinsic matrix, $\mathbf{W}$ the extrinsic matrix, and $\mathbf{J}$ the Jacobian matrix of the affine perspective projection. The Gaussian in image space is then:
$\mathcal{G}'(\mathbf{x}') = e^{-\frac{1}{2}(\mathbf{x}' - \mathbf{p}')^\top \mathbf{\Sigma}'^{-1} (\mathbf{x}' - \mathbf{p}')}$
, where $\mathbf{x}'$ is the pixel position transformed similarly to $\mathbf{p} \mapsto \mathbf{p}'$. Each 3D Gaussian primitive uses $\mathbf{c}$ and $\alpha$ to model appearance, with $\mathbf{c}$ representing view-dependent color (parameterized by spherical harmonics) and $\alpha$ the opacity. The pixel color $\mathbf{C}$ at $\mathbf{x}'$ is computed via volumetric alpha blending:
\begin{equation}
\label{eq4}
\mathbf{C}(\mathbf{x}') = \sum_{i=1}^N T_i \alpha_i \mathcal{G}'_i(\mathbf{x}') \mathbf{c}_i 
\quad
T_i = \prod_{j=1}^{i-1} (1 - \alpha_j \mathcal{G}'_i(\mathbf{x}'))
\end{equation}
where $\mathcal{G}'(\mathbf{x}')$ is the Gaussian with transformed $\mathbf{p}'$ and $\mathbf{\Sigma}'$, and $T_i$ is the transmittance along the ray.

To apply deformation to each Gaussian primitive, we apply $\mathbf{\phi}(\mathbf{X}, \mathbf{Z})$ to $\mathbf{p}$ and construct $\mathbf{\Sigma} = \mathbf{L}'\mathbf{L}'^\top$ with $\mathbf{L}' = \mathbf{F}(\mathbf{R} \mathbf{S})$. Here, $\mathbf{F} = \frac{\partial \mathbf{\phi}(\mathbf{p}, \mathbf{Z})}{\partial \mathbf{p}}$ is the deformation gradient, reflecting local deformation in continuum mechanics.

\section{Method}
\label{sec:method}

\begin{figure*}[htbp]
  \vspace{-5mm}
  \centering
  \includegraphics[width=\linewidth]{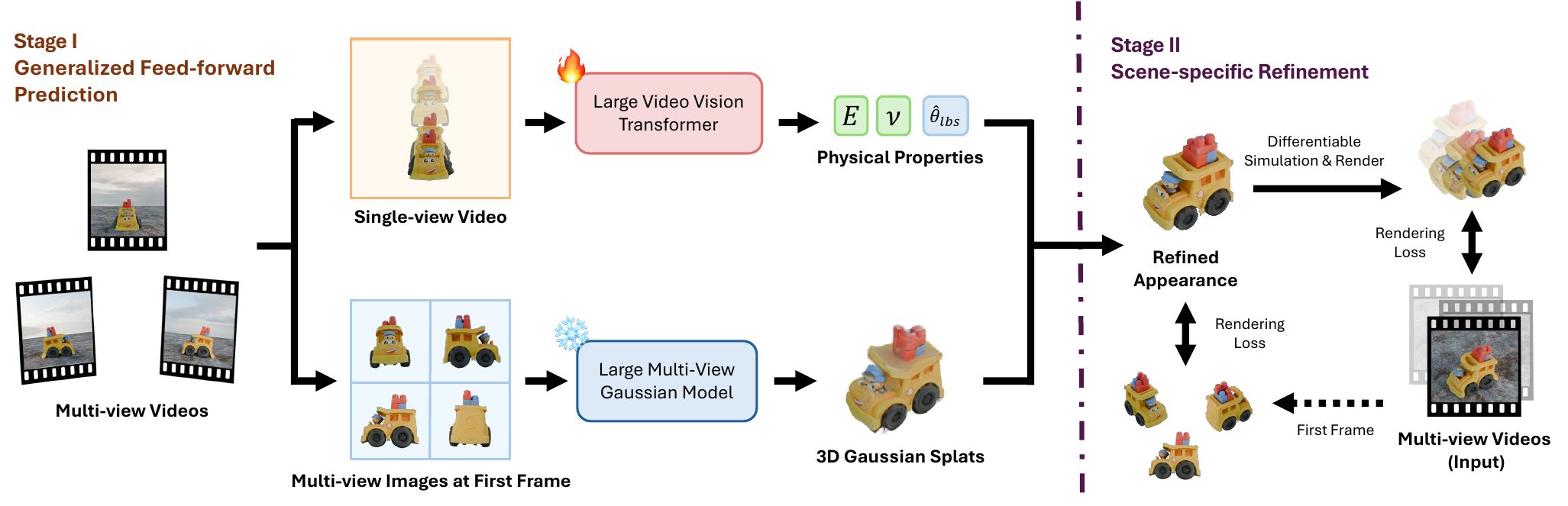}
  \vspace{-3mm}
  \caption{\textbf{An overview of \name, comprising two stages.} In Stage I, a generalizable feed-forward model reconstructs appearance, geometry, and physical properties, generating simulation-ready outputs. In Stage II, a lightweight optimization pipeline refines these estimated attributes to closely match the input video. We introduce a mesh-free reduced simulation based on Linear Blend Skinning (LBS), which provides high computational efficiency and versatile representational capability for high-fidelity dynamic reconstruction.
  \label{fig:pipeline}}
  \vspace{-3mm}
\end{figure*}

We aim to jointly reconstruct the appearance, geometry, and physical properties of the given target from posed multiview videos that describe the dynamics. We focus on elastic material modeled by the Neo-Hookean constitutive model to reduce the state space that our feed-forward predictor needs to learn, where we only predict Young’s modulus $E$, Poisson’s ratio $\nu$ and estimated scalar LBS weight $W_{\theta;j}(\mathbf{X}_i)$. Notably, our framework is not restricted to elastic materials and can be readily extended to various physical phenomena, which we demonstrate in the supplementary materials that our method generalizes across different material types. Our two-stage pipeline, illustrated in \cref{fig:pipeline}, is detailed below.

\subsection{Feed-forward Physical System
Identification}
\label{method:feedforward}

In the first stage, we develop several neural networks that learn physical world knowledge, enabling feed-forward reconstruction of the observed appearance, geometry, and physical configuration of the physical system from the video. 

We leverage the prior knowledge of physical dynamics by utilizing VideoMAE \cite{tong2022videomae} as the network backbone of our feed-forward predictor, which is a large video vision transformer pre-trained on a vast dataset of videos. The visual features extracted from the backbone are then decoded by several small MLPs, which function as the regression head to estimate physical properties. The whole network takes a single front-view video as input and regresses it to two physical parameters, $\{E, \nu\}$, relevant to elastic materials. Additionally, to enable mesh-free, reduced-order simulation, the network should also regress the LBS values $W_{\theta;j}(\mathbf{X}_i)$ used to deform positions for dynamics, as specified in \cref{eq:lbs}. However, as the LBS values are implicitly modeled using an MLP in~\cite{modi2024simplicits}, it becomes challenging to estimate them directly in a feed-forward manner.

To address this problem, we introduce a HyperNetwork~\cite{ha2016hypernetworks} approach for predicting the weights of MLP $\hat{\theta}_{\text{lbs}}$ for LBS estimation. This HyperNetwork is also implemented in a small MLP as a regression head, similar to the ones to predict $E$ and $\nu$. Additionally, it is tasked with regressing only the weights and biases of the final linear layer, keeping the other layers fixed at their default initialization. This design enhances the generalizability and robustness of LBS prediction during feed-forward inference. We demonstrate more details in our supplementary material. 

To recover geometry and appearance, we process the first multiview frames of the input videos by applying the pre-trained Large Multi-view Gaussian Model \cite{tang2025lgm}, which leverages the generalizable knowledge of the textured shape recovery trained with large-scale 3D datasets, and efficiently reconstruct them into 3D Gaussians as the shape representation, which is then normalized into a canonical space. 

Together, we recover the geometry, appearance, and physical properties through the two branches, as shown in ~\cref{method:feedforward}  Stage I, with a short inference time. This produces a simulation-ready prediction that meets all the requirements to be simulated with our simulation method. The feed-forward prediction is considered as a \textit{general} estimation, which is then further refined to more closely match the reference videos, resulting in a \textit{specific} estimation. More implementation details can be found in \cref{sec:implementation} and our supplementary material.

\subsection{Scene-specific Refinement}
\label{method:refinement} 

We conduct joint optimization of geometry, appearance, LBS, and physical parameters to better fit the reconstruction with the input multiview videos. Our lightweight optimization is significantly more efficient, completing in approximately 15 minutes, compared to existing methods that typically require around 1.5 hours. Detailed statistics are provided in~\cref{tab:efficiency}.

To improve the reconstruction quality of the shape and appearance, we first refine the 3D Gaussians via standard 3DGS training \cite{kerbl20233d}. Next, we refine the LBS estimation model to capture physical dynamics, enhancing its alignment with the specific dynamics of the given object. Usually, optimizing the LBS, as in Simplicits~\cite{modi2024simplicits}, requires precomputing the Jacobian of the deformation gradient with respect to transformations, $\mathbf{J}(\mathbf{X}) = \frac{\partial \mathbf{F}(\mathbf{X}, \mathbf{z})}{\partial \mathbf{z}}$, where $\mathbf{z}$ is the vector form of transformation $\mathbf{Z}$. Since $\mathbf{F} = \frac{\partial \phi(\mathbf{X}, \mathbf{z})}{\partial \mathbf{X}}$ includes only linear terms of $\mathbf{z}$, $\mathbf{J}$ depends solely on $\mathbf{X}$. For cubature points $C \subseteq \{\mathbf{X}_i \in \mathbb{R}^3 ~|~ i=1, 2, ..., n\}$, the Jacobian $\mathbf{J} \in \mathbb{R}^{9N_c \times m \times m}$ grows large with increasing cubature points $N_c$ and handles $m$, necessitating computation through auto-differentiation. Precomputing this Jacobian is manageable if done once for fixed neural LBS, but further LBS optimization makes this cost-prohibitive.

In our method, we accelerate the refinement (and simulation) by introducing a Neural Jacobian module. 

\noindent \textbf{Neural Jacobian.} We employ a neural network trained to predict $\mathbf{J}_{\theta}(\mathbf{X})$ instead of computing it explicitly.
The Neural Jacobian is trained following the LBS training using the loss function below

\vspace{-4mm}
\begin{equation}
\mathcal{L}_{J}=||\mathbf{J}_\theta(\mathbf{X})\mathbf{z}+\mathbf{I}-\mathbf{F}(\mathbf{X},\mathbf{z})||_1,
\end{equation}

\noindent where $\mathbf{J}_\theta(\mathbf{X})\mathbf{z}+\mathbf{I}$ is an estimation of the deformation gradient $\mathbf{F}(\mathbf{X},\mathbf{z})$ and its ground truth is much cheaper to get via finite differences. The training samples for $\mathbf{X}$ and $\mathbf{z}$ are generated in a data-free manner the same as~\cite{modi2024simplicits}.We validate the effectiveness of the Neural Jacobian in~\cref{sec:ablation} and our supplementary material. The speed-up is shown in \cref{tab:efficiency}.

Then, we optimize the physical parameters, with fine-tuning the LBS and the corresponding Neural Jacibian at the same time, to match the input videos. We use rendering loss to supervise the optimization. This process can be formulated as:

\begin{equation}
\begin{aligned}
\theta_{lbs}^*, \theta_{jac}^*, E^*, \nu^* &= \mathop{\rm{argmin}}\limits_{\theta_{lbs}, \theta_{jac}, E, \nu} \mathcal{L}_{rendering} \\
\mathcal{L}_{rendering}=\frac{1}{N \Delta s}\sum_{i=1}^N &\sum_{t=s}^{s+\Delta s} \| \mathbf{C}_{pred}(i, t) - \mathbf{C}_{gt}(i, t) \|_2^2.
\end{aligned}
\end{equation}
Here, $\mathbf{C}_{\text{pred}}$ represents the rendering sequence from the simulation steps $\{\mathbf{z}_s, \mathbf{z}_{s+1}, \dots, \mathbf{z}_{s+\Delta s}\}$, $\mathbf{C}_{\text{gt}}$ is the reference rendering sequence, and $N$ denotes the number of views.
For efficiency, we set $\Delta s = 4$ and randomly sample $s$ from $s'$ to $T - \Delta s$ in each iteration, in which $s'$ is the first frame where $\mathbf{C}_{\text{pred}}$ is different from $\mathbf{C}_{\text{gt}}$. This allows the process to cover the entire valid observation.
\section{Implementation Details}
\label{sec:implementation}

\subsection{Feed-forward Physical System Identification}
\paragraph{Dataset.}
We choose 50k high-quality 3D objects from Objaverse \cite{deitke2023objaverse} to construct our dataset (49k for training and 1k for validation). For each object, we generate an animation with the motion of falling to the ground at 448 $\times$ 448 resolution simulated by our reduced simulator, with randomly sampled $E \in [10^4, 10^6]$, $\nu \in [0.2, 0.5]$.

\vspace{-3mm}
\paragraph{Implementation.}
We use two identical 4-layer MLPs to predict the scalar $E$ and $\nu$ and a 4-layer MLP as the hypernetwork to predict the final linear layer of the LBS network. We trained the whole network on one NVIDIA-L40 GPU for 1 day with the Adam~\cite{kingma2014adam} and a learning rate of $10^{-5}$, where the backbone's weights are fine-tuned from pretraining and the regression heads are trained from scratch.

\subsection{Physical System Refinement}
\paragraph{Dataset.}
To evaluate the performance of our full pipeline, we use both a synthetic dataset and a real-world dataset. 

The synthetic dataset is a mesh dataset that contains 12 delicate objects collected from Google Scanned Objects (GSO) \cite{downs2022google} with complex geometry and detailed texture. We use FEM to simulate animations in the most accurate physic as references. We rendered each animation from 12 different viewpoints at 448 $\times$ 448 resolution for 24 frames. The first 16 frames are treated as observation, and the 8 frames remaining are references for future state prediction. 

For the real-world dataset, we captured 3 different animations (See \cref{fig:real_world}) \textit{orange}, \textit{bird} and \textit{cup} with four posed cameras at surrounding views. We use BackgroundMattingV2 \cite{Lin_2021} with post-processing to obtain the mask of the object.

\vspace{-3mm}
\paragraph{Implementation.}
We first refine the 3D Gaussians following the original 3DGS \cite{kerbl20233d} and use the data-free method from \cite{modi2024simplicits} to train the full LBS layers and the corresponding Neural Jacobian. Afterwards, we jointly optimize $\{\theta_{lbs}, \theta_{jac}, E, \nu\}$ for 400 iterations. We also use the Adam optimizer and the learning rates are set to $\{5 \times 10^{-7}, 5 \times 10^{-7}, 5 \times 10^{-3}, 1 \times 10^{-3} \}$. We use 10 control handles and 500 cubature points for simulation. We use Farthest Point Sampling (FPS) to sample cubature points.
\section{Experiments}
\label{sec:experiments}
\subsection{Baselines and Metrics}
We compare our method with the state-of-the-art methods: GIC~\cite{cai2024gaussian}, Spring-GS~\cite{zhong2025reconstruction}, and PAC-NeRF~\cite{li2023pac} on the dynamic reconstruction task and the future state prediction task at both synthetic and real-world datasets. We use the Peak Signal-to-Noise Ratio (PSNR), Structural Similarity Index Metric (SSIM), and video perceptual loss (FoVVDP) as the metrics for evaluation. We additionally report the running time of each method to assess runtime efficiency in \cref{tab:efficiency}.

\begin{figure*}[t]
  \vspace{-5mm}
  \centering
  \includegraphics[width=\linewidth]{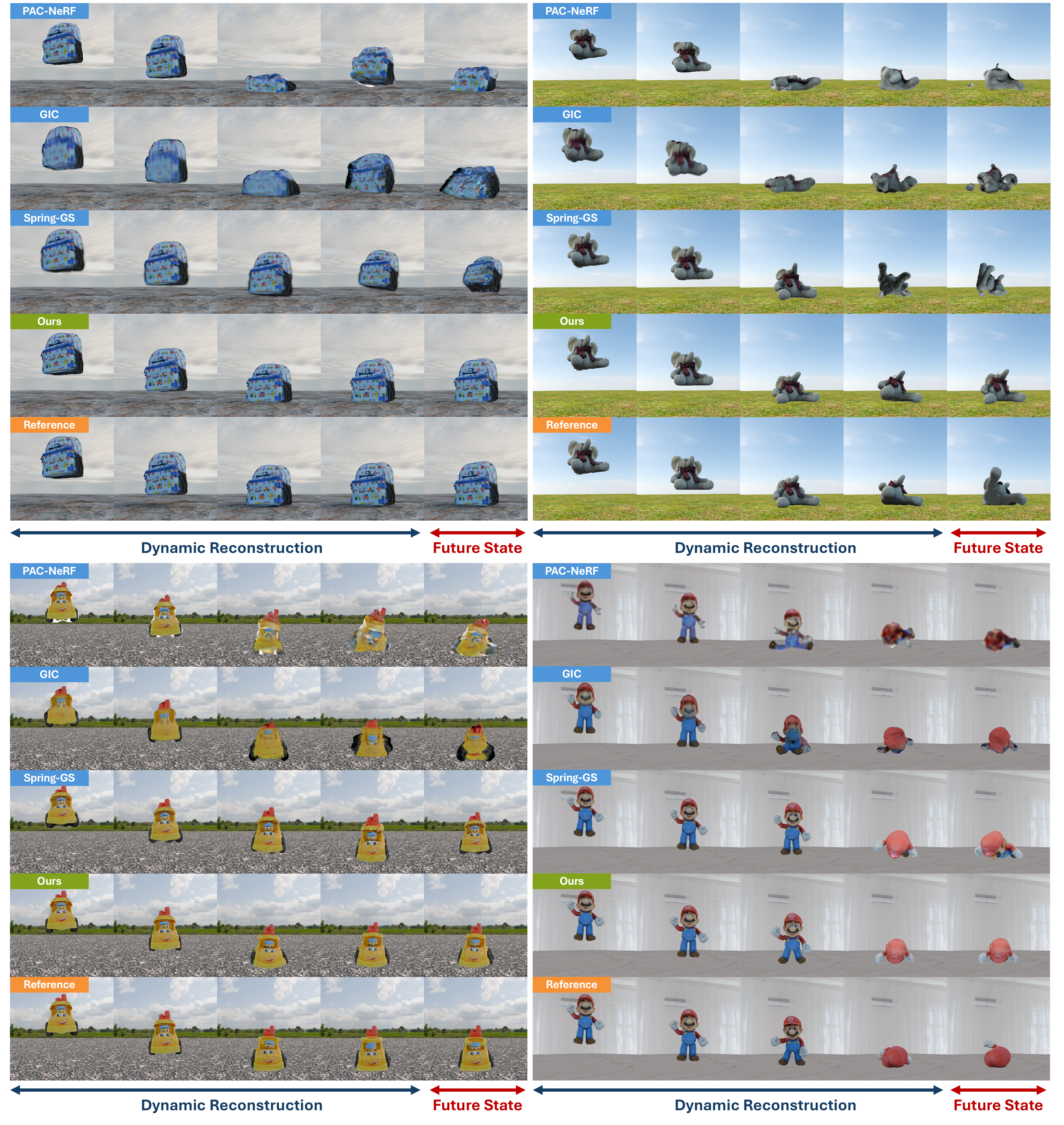}
  \vspace{-5mm}
  \caption{\label{fig:synthetic} Comparison with the SOTA methods~\cite{li2023pac, cai2024gaussian, zhong2025reconstruction} on physics-aware dynamic reconstruction from multi-view videos (reference). Our method achieves the best quality in terms of textured shape and physical dynamics.}
  \vspace{-5mm}
\end{figure*}

\subsection{Evaluation on the synthetic dataset}
Following previous methods~\cite{li2023pac, zhong2025reconstruction, kaneko2024improving, cai2024gaussian}, we evaluate our method and baselines for dynamic reconstruction on the 12 diverse synthetic test cases. Both qualitative results (\cref{fig:synthetic}) and quantitative results (\cref{tab:synthetic}) show that our method \name achieves a much higher quality of reconstruction for appearance and physics compared with all the SOTA methods across different objects. To be more specific, previous methods rely on optimizing dynamic NeRF or 3D Gaussians to model appearance, a process that is challenging in high-dimensional spaces and often results in blurred textures as shown in \cref{fig:synthetic}. In contrast, our pipeline enables explicit deformation guided by a deformation field based on 3D Gaussians, preserving high-quality details optimized in the canonical space. Furthermore, baseline models are constrained to a differentiable simulator with a symplectic solver, which introduces oscillations and instability, compromising the realism of the simulations. Unlike these models, our implicit solver within the differentiable simulator provides a more accurate and efficient simulation.

\begin{table*}[t]
\vspace{-3mm}
\centering
\caption{\label{tab:synthetic} Quantitative Comparison with Previous Methods in Dynamic Reconstruction.}
\vspace{-3mm}
\resizebox{\linewidth}{!}{
\begin{tabular}{c|c|cccccccccccc|c}
\toprule
\multicolumn{2}{c|}{} & backpack & bell & blocks & bus & cream & elephant & grandpa & leather & lion & mario & sofa & turtle & Mean \\ 
\midrule
\multirow{4}{*}{\rotatebox[origin=c]{90}{PSNR $\uparrow$}} & PAC-NeRF & 19.37 & 25.00 & 23.36 & 20.72 & 23.24 & 22.27 & 21.63 & 20.85 & 22.66 & 21.01 & 22.49 & 22.19 & 22.06 \\ 
& Spring-Gaus & 17.42 & 20.49 & 22.78 & 20.06 & 23.58 & 21.30 & 21.64 & 18.29 & 21.95 & 20.23 & 21.89 & 21.23 & 20.91 \\ 
& GIC & 18.94 & 19.55 & 18.78 & 20.84 & 23.81 & 21.86 & 20.50 & 21.00 & 19.33 & 21.28 & 24.16 & 22.09 & 21.01 \\ 
& Ours (full) & \textbf{28.30} & \textbf{30.14} & \textbf{33.49} & \textbf{27.76} & \textbf{35.75} & \textbf{29.29} & \textbf{27.52} & \textbf{32.55} & \textbf{28.27} & \textbf{27.35} & \textbf{30.44} & \textbf{31.13} & \textbf{30.17} \\ 
\midrule
\multirow{4}{*}{\rotatebox[origin=c]{90}{SSIM $\uparrow$}} & PAC-NeRF & 0.887 & 0.956 & 0.940 & 0.908 & 0.893 & 0.922 & 0.939 & 0.932 & 0.936 & 0.921 & 0.926 & 0.923 & 0.924 \\ 
& Spring-Gaus & 0.867 & 0.941 & 0.941 & 0.903 & 0.912 & 0.919 & 0.948 & 0.917 & 0.937 & 0.920 & 0.921 & 0.920 & 0.920 \\ 
& GIC & 0.903 & 0.945 & 0.930 & 0.925 & 0.922 & 0.936 & 0.948 & 0.949 & 0.934 & 0.938 & 0.942 & 0.936 & 0.934 \\ 
& Ours (full) & \textbf{0.944} & \textbf{0.972} & \textbf{0.978} & \textbf{0.944} & \textbf{0.966} & \textbf{0.955} & \textbf{0.962} & \textbf{0.977} & \textbf{0.957} & \textbf{0.949} & \textbf{0.954} & \textbf{0.969} & \textbf{0.961} \\
\midrule
\multirow{4}{*}{\rotatebox[origin=c]{90}{FoVVDP $\uparrow$}} & PAC-NeRF & 6.043 & 7.473 & 7.001 & 6.540 & 5.991 & 6.791 & 6.626 & 6.485 & 7.006 & 6.876 & 6.543 & 6.711 & 6.674 \\ 
& Spring-Gaus & 5.455 & 6.862 & 6.890 & 6.377 & 5.899 & 6.524 & 6.998 & 5.988 & 6.902 & 6.153 & 6.300 & 6.569 & 6.410 \\ 
& GIC & 6.130 & 6.230 & 6.062 & 6.552 & 5.889 & 6.907 & 6.855 & 6.737 & 6.331 & 6.985 & 7.069 & 6.782 & 6.544 \\ 
& Ours (full) & \textbf{8.341} & \textbf{8.288} & \textbf{8.943} & \textbf{7.948} & \textbf{9.181} & \textbf{8.307} & \textbf{7.830} & \textbf{9.007} & \textbf{7.866} & \textbf{7.771} & \textbf{8.049} & \textbf{8.820} & \textbf{8.363} \\
\bottomrule
\end{tabular}}
 \vspace{-3mm}
\end{table*}

\subsection{Ablation Study}
\label{sec:ablation}
We conduct extensive ablation studies on our key designs. \cref{tab:ablation} summarize the quantitative results. Since only 4 views are used in LGM~\cite{tang2025lgm} in our Stage I, it is difficult to reconstruct the accurate appearance and geometry at inference time, resulting in compromised quantitative results ({\textit{Ours (Stage I only)}}). Nevertheless, the predicted physical properties from Stage I are effective enough to produce high-quality simulations. This is validated by\textit{Ours (Stage I+refine GS)}, where we solely refine the 3D Gaussians from LGM initialization without changing any physical properties. This demonstrates that appearance and geometry are critical for the overall dynamic reconstruction. \textit{Ours (Stage I+fit GS) is a similar ablation where the 3D Gaussians are trained from scratch, demonstrating a worse result than using LGM prediction as initialization}. \textit{Ours (full w/o fine-tune LBS)} shows a further improvement when adding the optimization of the $E$ and $\nu$, and our full model that unlocks the LBS reaches the best. Additionally, \textit{Ours (full w/o Stage I Phys.)} shows purely optimization results with random physics initialization, for which we ran the experiments $3$ times with random samples of $E \in [10^4, 10^6]$, $\nu \in [0.2, 0.5]$, same as the prediction range of our feed-forward predictor. This result suggests that a reliable initialization is crucial for achieving final convergence.

\begin{table*}[t]
\centering
\caption{\label{tab:ablation} Ablation of dynamic reconstruction.}
\vspace{-3mm}
\resizebox{\linewidth}{!}{
\begin{tabular}{c|c|cccccccccccc|c}
\toprule
\multicolumn{2}{c|}{} & backpack & bell & blocks & bus & cream & elephant & grandpa & leather & lion & mario & sofa & turtle & Mean \\ 
\midrule
\multirow{4}{*}{\rotatebox[origin=c]{90}{PSNR $\uparrow$}}& Ours (Stage I only) & 19.54 & 21.34 & 20.50 & 19.62 & 18.08 & 19.89 & 19.46 & 15.53 & 21.36 & 19.07 & 20.48 & 22.36 & 19.77 \\ 
& Ours (Stage I+fit GS) & 26.87 & 26.59 & 32.57 & 26.53 & 34.82 & 26.99 & 24.45 & 31.21 & \cellcolor{LightYellow1}27.62 & 24.83 & 30.01 & 30.33 & 28.57 \\ 
& Ours (Stage I+refine GS) & 26.52 & 27.61 & 32.49 & 26.63 & 34.54 & 27.05 & 24.42 & \cellcolor{LightYellow1}31.32 & 27.54 & 25.73 & \cellcolor{Goldenrod}30.11 & 30.37 & 28.69 \\ 
& Ours (full w/o fine-tune LBS) & \cellcolor{LightYellow1}27.37 & \cellcolor{LightYellow1}27.97 & \cellcolor{LightYellow1}32.98 & \cellcolor{LightYellow1}27.01 & \cellcolor{LightYellow1}35.31 & \cellcolor{LightYellow1}27.93 & \cellcolor{Goldenrod}27.34 & 31.06 & \cellcolor{YellowOrange}28.27 & \cellcolor{Goldenrod}26.60 & 29.90 & \cellcolor{LightYellow1}31.00 & \cellcolor{Goldenrod}29.40 \\ 
& Ours (full w/o Stage I Phys.)  & \cellcolor{Goldenrod}27.63 & \cellcolor{YellowOrange}30.79 & \cellcolor{Goldenrod}33.02 & \cellcolor{YellowOrange}28.35 & \cellcolor{YellowOrange}35.75 & \cellcolor{Goldenrod}28.59 & \cellcolor{LightYellow1}26.72 & \cellcolor{YellowOrange}32.71 & 27.06 & \cellcolor{LightYellow1}26.58 & \cellcolor{LightYellow1}30.09 & \cellcolor{YellowOrange}31.16 & \cellcolor{LightYellow1}29.87 \\ 
& Ours (full) & \cellcolor{YellowOrange}28.30 & \cellcolor{Goldenrod}30.14 & \cellcolor{YellowOrange}33.49 & \cellcolor{Goldenrod}27.76 & \cellcolor{Goldenrod}35.75 & \cellcolor{YellowOrange}29.29 & \cellcolor{YellowOrange}27.52 & \cellcolor{Goldenrod}32.55 & \cellcolor{YellowOrange}28.27 & \cellcolor{YellowOrange}27.35 & \cellcolor{YellowOrange}30.44 & \cellcolor{Goldenrod}31.13 & \cellcolor{YellowOrange}30.17 \\
\midrule
\multirow{4}{*}{\rotatebox[origin=c]{90}{SSIM $\uparrow$}} & Ours (Stage I only) & 0.883 & 0.950 & 0.924 & 0.897 & 0.909 & 0.904 & 0.938 & 0.895 & 0.931 & 0.917 & 0.913 & 0.921 & 0.915 \\ 
& Ours (Stage I+fit GS) & 0.931 & 0.959 & 0.973 & 0.937 & 0.955 & 0.942 & 0.944 & 0.973 & \cellcolor{LightYellow1}0.954 & 0.936 & 0.952 & 0.963 & 0.952 \\ 
& Ours (Stage I+refine GS) & 0.929 & 0.960 & 0.974 & 0.937 & 0.953 & 0.943 & 0.945 & 0.974 & 0.954 & 0.942 & \cellcolor{LightYellow1}0.952 & 0.962 & 0.952 \\ 
& Ours (full w/o fine-tune LBS) & \cellcolor{LightYellow1}0.934 & \cellcolor{LightYellow1}0.965 & \cellcolor{Goldenrod}0.975 & \cellcolor{LightYellow1}0.942 & \cellcolor{LightYellow1}0.960 & \cellcolor{LightYellow1}0.947 & \cellcolor{LightYellow1}0.961 & \cellcolor{LightYellow1}0.974 & \cellcolor{YellowOrange}0.957 & \cellcolor{LightYellow1}0.946 & 0.951 & \cellcolor{LightYellow1}0.968 & \cellcolor{Goldenrod}0.957 \\ 
& Ours (full w/o Stage I Phys.)  & \cellcolor{Goldenrod}0.938 & \cellcolor{YellowOrange}0.973 & \cellcolor{LightYellow1}0.975 & \cellcolor{YellowOrange}0.948 & \cellcolor{Goldenrod}0.962 & \cellcolor{Goldenrod}0.952 & \cellcolor{Goldenrod}0.961 & \cellcolor{Goldenrod}0.977 & 0.952 & \cellcolor{YellowOrange}0.951 & \cellcolor{Goldenrod}0.953 & \cellcolor{YellowOrange}0.969 & \cellcolor{LightYellow1}0.959 \\ 
& Ours (full) & \cellcolor{YellowOrange}0.944 & \cellcolor{Goldenrod}0.972 & \cellcolor{YellowOrange}0.978 & \cellcolor{Goldenrod}0.944 & \cellcolor{YellowOrange}0.966 & \cellcolor{YellowOrange}0.955 & \cellcolor{YellowOrange}0.962 & \cellcolor{YellowOrange}0.977 & \cellcolor{YellowOrange}0.957 & \cellcolor{Goldenrod}0.949 & \cellcolor{YellowOrange}0.954 & \cellcolor{Goldenrod}0.969 & \cellcolor{YellowOrange}0.961 \\
\midrule
\multirow{4}{*}{\rotatebox[origin=c]{90}{FoVVDP $\uparrow$}} & Ours (Stage I only) & 6.616 & 6.175 & 6.481 & 6.545 & 4.450 & 6.188 & 6.064 & 5.661 & 6.834 & 5.912 & 6.062 & 7.330 & 6.193 \\ 
& Ours (Stage I+fit GS) & 7.778 & 6.884 & 8.792 & 7.622 & 9.117 & 7.468 & 6.246 & 8.949 & \cellcolor{LightYellow1}7.681 & 6.325 & \cellcolor{LightYellow1}8.034 & 8.628 & 7.794 \\ 
& Ours (Stage I+refine GS) & 7.664 & 7.000 & 8.802 & 7.657 & 9.105 & 7.473 & 6.249 & \cellcolor{LightYellow1}8.968 & 7.675 & 6.825 & \cellcolor{YellowOrange}8.060 & 8.632 & 7.843 \\ 
& Ours (full w/o fine-tune LBS) & \cellcolor{Goldenrod}8.080 & \cellcolor{LightYellow1}7.680 & \cellcolor{Goldenrod}8.873 & \cellcolor{LightYellow1}7.770 & \cellcolor{YellowOrange}9.199 & \cellcolor{LightYellow1}7.916 & \cellcolor{Goldenrod}7.676 & 8.929 & \cellcolor{YellowOrange}7.866 & \cellcolor{Goldenrod}7.312 & 7.994 & \cellcolor{LightYellow1}8.729 & \cellcolor{Goldenrod}8.169 \\ 
& Ours (full w/o Stage I Phys.)  & \cellcolor{LightYellow1}7.996 & \cellcolor{YellowOrange}8.439 & \cellcolor{LightYellow1}8.861 & \cellcolor{YellowOrange}8.107 & \cellcolor{LightYellow1}9.158 & \cellcolor{Goldenrod}8.064 & \cellcolor{LightYellow1}7.433 & \cellcolor{YellowOrange}9.011 & 7.438 & \cellcolor{LightYellow1}7.218 & 7.942 & \cellcolor{Goldenrod}8.782 & \cellcolor{LightYellow1}8.204 \\ 
& Ours (full) & \cellcolor{YellowOrange}8.341 & \cellcolor{Goldenrod}8.288 & \cellcolor{YellowOrange}8.943 & \cellcolor{Goldenrod}7.948 & \cellcolor{Goldenrod}9.181 & \cellcolor{YellowOrange}8.307 & \cellcolor{YellowOrange}7.830 & \cellcolor{Goldenrod}9.007 & \cellcolor{YellowOrange}7.866 & \cellcolor{YellowOrange}7.771 & \cellcolor{Goldenrod}8.049 & \cellcolor{YellowOrange}8.820 & \cellcolor{YellowOrange}8.363 \\
\bottomrule
\end{tabular}}
\vspace{-3mm}
\end{table*}

\subsection{Future State Prediction}
Like Spring-Gaus \cite{zhong2025reconstruction} and GIC \cite{cai2024gaussian}, we also perform a test of future state prediction to evaluate how our model's simulation aligns the observed videos in future frames. We report an average result across all the test cases on our synthetic dataset in~\cref{tab:future}, for which both our method and baselines predict 8 frames after reconstructing from 16 frames. The results show that our method keeps better accuracy than all the baselines.

\subsection{Evaluation on the real-world dataset}
\begin{figure}[htbp]
  \centering
  \includegraphics[width=\linewidth]{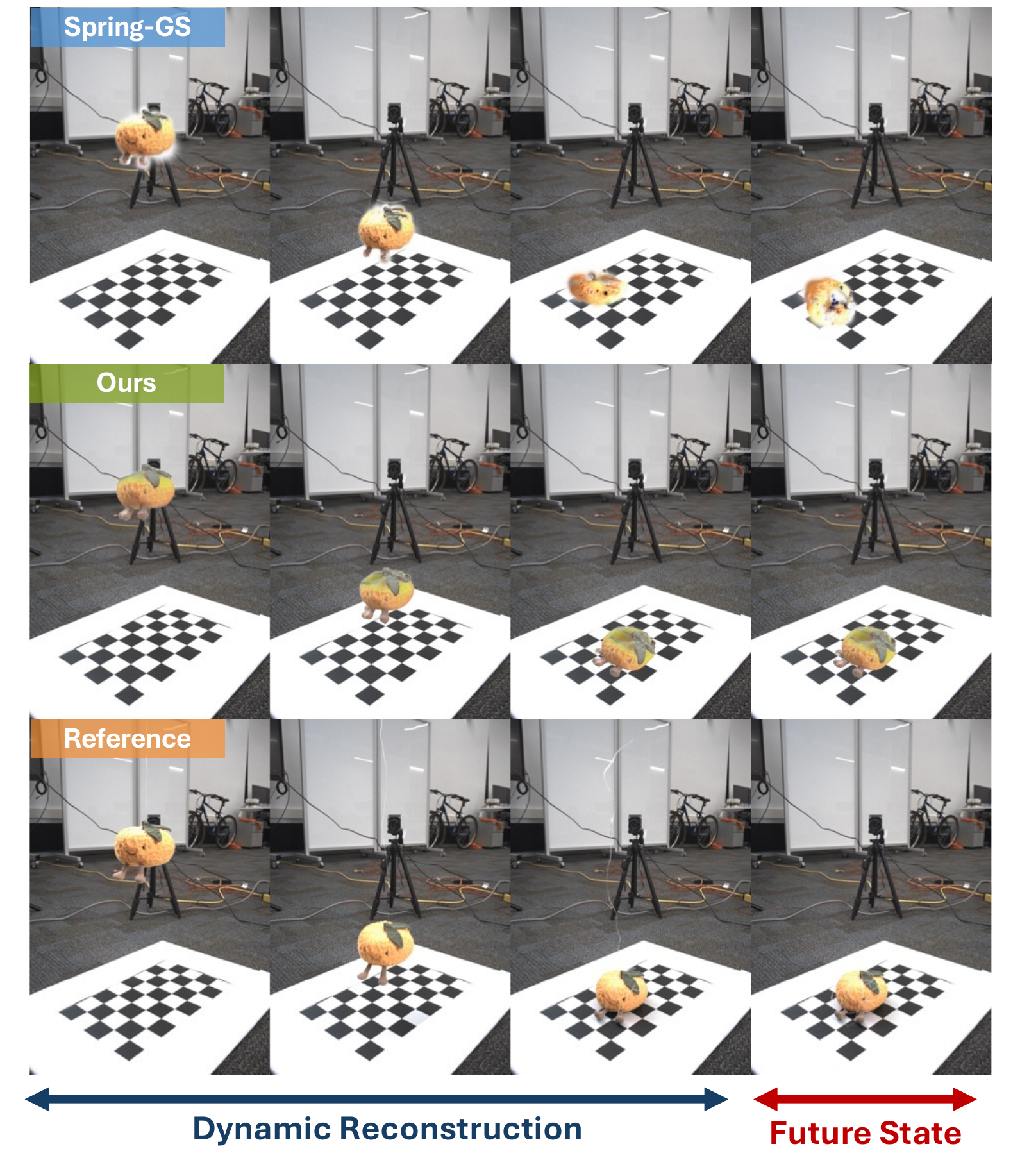}
  \vspace{-5mm}
  \caption{Visualization of dynamic reconstruction results of \name on the real-world object.\label{fig:real_world}}
  \vspace{-3mm}
\end{figure}

We next evaluate our model on the real-world dataset. Obtaining accurate 3D Gaussian representations from sparse viewpoints in our real-world dataset poses a significant challenge. To address this issue, we employ the registration network introduced by Spring-Gaus \cite{zhong2025reconstruction} to align the poses of the 3D Gaussians estimated by LGM \cite{tang2025lgm} in Stage I with the real-world camera poses. Our approach then leverages these registered static 3D Gaussians, in the manner of Spring-Gaus, to facilitate reconstruction and simulation. We compare our method with Spring-Gaus for both dynamic reconstruction and future state prediction, as shown in \cref{fig:real_world} and \cref{tab:real_world}. Our approach demonstrates enhanced capability in modeling real-world objects, particularly in future state prediction.

\begin{table}[t]
\vspace{-2mm}
\centering
\caption{\label{tab:real_world} Evaluation on the real-world object.}
\vspace{-3mm}
\resizebox{0.8\linewidth}{!}{
\begin{tabular}{c|c|ccc|c}
\toprule
\multicolumn{2}{c|}{} & orange & bird & cup & Mean \\ 
\midrule
\multirow{2}{*}{\rotatebox[origin=c]{90}{\scriptsize PSNR $\uparrow$}}
& Spring-Gaus & 28.69 & 25.08 & 24.39 & 26.05 \\ 
& Ours (full) & \textbf{30.11} & \textbf{26.02} & \textbf{25.24} & \textbf{27.12} \\
\midrule
\multirow{2}{*}{\rotatebox[origin=c]{90}{\scriptsize SSIM $\uparrow$}} 
& Spring-Gaus & 0.987 & 0.980 & 0.979 & 0.982 \\ 
& Ours (full) & 0.987 & \textbf{0.981} & \textbf{0.980} & \textbf{0.983} \\
\midrule
\multirow{2}{*}{\rotatebox[origin=c]{90}{\tiny FoVVDP$\uparrow$}} 
& Spring-Gaus & 8.379 & 7.494 & 7.447 & 7.773 \\ 
& Ours (full) & \textbf{8.623} & \textbf{7.554} & 7.447 & \textbf{7.875} \\
\bottomrule
\end{tabular}}
\vspace{-3mm}
\end{table}

\begin{table}[t]
\centering
\caption{\label{tab:future} Comparison on future state prediction.}
\vspace{-2mm}
\resizebox{0.8\linewidth}{!}{
\begin{tabular}{c|ccc}
\toprule
& PSNR $\uparrow$ & SSIM $\uparrow$ & FoVVDP $\uparrow$ \\ 
\midrule
PAC-NeRF & 20.11 & 0.913 & 5.948 \\ 
Spring-Gaus & 18.32 & 0.905 & 5.443 \\ 
GIC & 19.20 & 0.916 & 5.702 \\ 
Ours (full) & \textbf{25.07} & \textbf{0.945} & \textbf{7.770} \\
\bottomrule
\end{tabular}}
\vspace{-5mm}
\end{table}

\subsection{Comparison of Efficiency}
Though using an implicit Euler solver with Newton's method and line search, our method is still much more efficient regarding differentiable simulation. This is because of four reasons: \textbf{(1)} The implicit Euler solver requires fewer time steps; \textbf{(2)} The simulation and optimization is operated on a reduced dimension; \textbf{(3)} We design a neural Jacobian for faster precomputation and \textbf{(4)} Our strategy of using partial frames. 

We compare the computation time among our method and baselines for one optimization iteration that contains one forward and backward pass (consider using all 12 views on \textit{backpack} case). We also report the whole training time for all the methods with the default settings. Our results in \cref{tab:efficiency} show that our method is even faster than the efficient Spring-Gaus method, and our proposed neural Jacobian saves more time when using more cubature points and handles in simulation. All the performances are tested on one NVIDIA-RTX-4090 GPU.

\begin{table}[t]
\vspace{-2mm}
\centering
\caption{\label{tab:efficiency} Comparison with existing methods on runtime performance. The results in $(\cdot)$ is the case that uses 40 handles and 2000 cubature points for more accurate simulation.}
\vspace{-3mm}
\resizebox{0.95\linewidth}{!}{
\begin{tabular}{c|c|c}
\toprule    
& Per Iteration Time & Total Training Time  \\ 
\midrule
GIC & 37.33s & 120min \\
PAC-NeRF & 29.04s & 84min \\
Spring-GS & \cellcolor{LightYellow1} 8.08s & \cellcolor{LightYellow1} 54min \\
Ours (w/o $J_\theta$)  & \cellcolor{Goldenrod} 3.22s~(13.11s) & \cellcolor{Goldenrod} 26 min \\
Ours (full) & \cellcolor{YellowOrange} 1.44s (2.11s) & \cellcolor{YellowOrange} 15 min \\
\bottomrule
\end{tabular}}
\vspace{-5mm}
\end{table}


\vspace{-2mm}
\section{Conclusion}
\vspace{-2mm}
\label{sec:conclusion}
In this paper, we present \name, a novel and robust framework for high-fidelity and generalizable reconstruction of textured shapes and physical properties directly from video data. Our approach overcomes key limitations in existing methods by incorporating a feed-forward model that efficiently provides generalizable initial estimation, alongside a differentiable, reduced-order simulator utilizing Linear Blend Skinning for fast and precise optimization of appearance, geometry, and physical properties. After the reconstruction, \name enables high-quality, mesh-free simulation with high efficiency. Comprehensive experiments demonstrate that \name achieves state-of-the-art performance in both accuracy and efficiency, representing a significant advancement in video-based system identification.

\vspace{-2mm}
\section{Limitation and Future Work}
\label{sec:limitation}
Our approach is limited in reconstructing and simulating complex materials, e.g. fluid, since we use a reduced-order simulation method. Future works include further enhancing the ability to express more complex material and motions. Another direction is to merge the two branches of our Stage I and train a unified feed-forward network to predict 3D Gaussians together with point-wise physical properties.

{
    \small
    \bibliographystyle{ieeenat_fullname}
    \bibliography{main}
}

\clearpage
\setcounter{page}{1}
\maketitlesupplementary

This supplementary material covers the following sections:  More Implementation Details~(\cref{sec:details}); More Results on Dynamic Reconstruction (\cref{sec:dy}); Generalization Capability~(\cref{sec:gen}). Please refer to our supplementary video for a more comprehensive overview,

\section{More Implementation Details}
\label{sec:details}

\subsection{Large Video Vision Transformer}
\label{sec:ff}

\begin{figure}[!h]
  \centering
  \includegraphics[width=\linewidth]{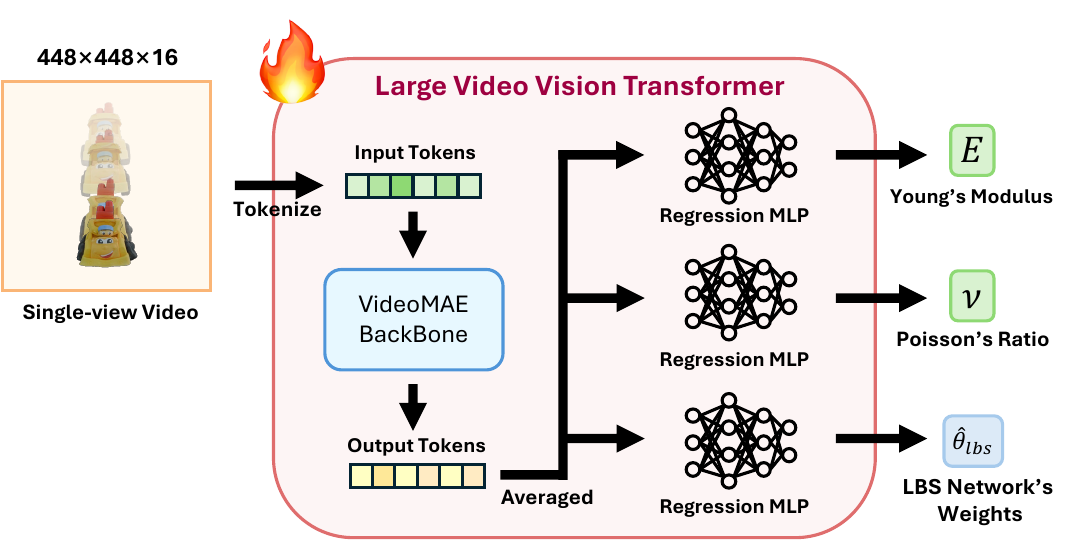}
  \caption{\label{fig:ff} Detailed pipeline of the large video vision transformer.}
  \vspace{-2mm}
\end{figure}

The pipeline of our Large Video Vision Transformer is shown in \cref{fig:ff}. In our framework, we fine-tune the backbone network, VideoMAE \cite{tong2022videomae}, which is pre-trained on 16-frame videos at a resolution of \(224 \times 224\). To adapt it to a higher resolution (\(448 \times 448\) in our setting), we interpolate the pre-trained positional embeddings to align with the updated number of input tokens. The output tokens are averaged across all the patches before being sent into the regression MLPs. The regression MLPs for predicting $E$ and $\nu$ are identical and with widths of $[768, 512, 256, 128, 1]$. The regression MLP for predicting $\hat \theta_{lbs}$ has widths of $[768, 650, 650, 650, 650]$ where the width of the last layer is equal to the number of trainable parameters for a linear layer. We demonstrate in~\cref{tab:LBS} that it is better to predict only the last layer of $\hat \theta_{lbs}$ and keep the first 7 layers fixed for consistency with the optimization stage (Stage II) than to predict full layers in our task. This is because Hypernetwork predicts $\sim$ 30k network parameters for full-layer LBS, making training much more difficult than our one-layer prediction design. We use GELU~\cite{hendrycks2016gaussian} as the activation function for all regression MLPs.

\begin{table}[!h]
\centering
\resizebox{\linewidth}{!}{
\begin{tabular}{c|ccc}
\toprule
& PSNR $\uparrow$ & SSIM $\uparrow$ & FoVVDP $\uparrow$ \\ 
\midrule
One-layer prediction (Ours) & \textbf{28.83±3.06} & \textbf{0.954±0.014} & \textbf{7.907±0.859} \\ 
Full-layer prediction & 28.53±3.21 & 0.953±0.015 & 7.782±0.886 \\ 
\midrule
GT (data-free train) & 29.40±2.59 & 0.957±0.012 & 8.169±0.579 \\
\bottomrule
\end{tabular}}
\caption{\label{tab:LBS}Quantitative results in Dynamic Reconstruction across different LBS prediction settings using the same optimized geometry and physical parameters for fairness.}
\end{table}

\begin{table}[h]
\centering
\resizebox{\linewidth}{!}{
\begin{tabular}{c|ccc}
\toprule
& Forward Time & Backward Time & $\frac{1}{N}\Sigma_{i=1}^N||\mathbf{J}_{\theta}^i-\mathbf{J}_{gt}^i||_2^2$ \\ 
\midrule
4 blocks w/o PE & 0.00081s & 0.00158s & $9.22 \times 10^{-7}$ \\ 
2 blocks & \textbf{0.00052s} & \textbf{0.00099s} & $5.12 \times 10^{-7}$ \\ 
4 blocks (Ours) & 0.00086s & 0.00166s & $\boldsymbol{4.04 \times 10^{-8}}$ \\ 
\midrule
    GT Jacobian & 6.51398s & 0.00014s & - \\ 
\bottomrule
\end{tabular}}
\caption{\label{tab:nj}Speed (time per iteration) and average accuracy across different Neural Jacobian models. All the values are tested under the setting of 2000 points \& 10 handles on one NVIDIA-RTX-4090 GPU.}
\end{table}

\begin{figure*}[!h]
  \centering
  \includegraphics[width=\linewidth]{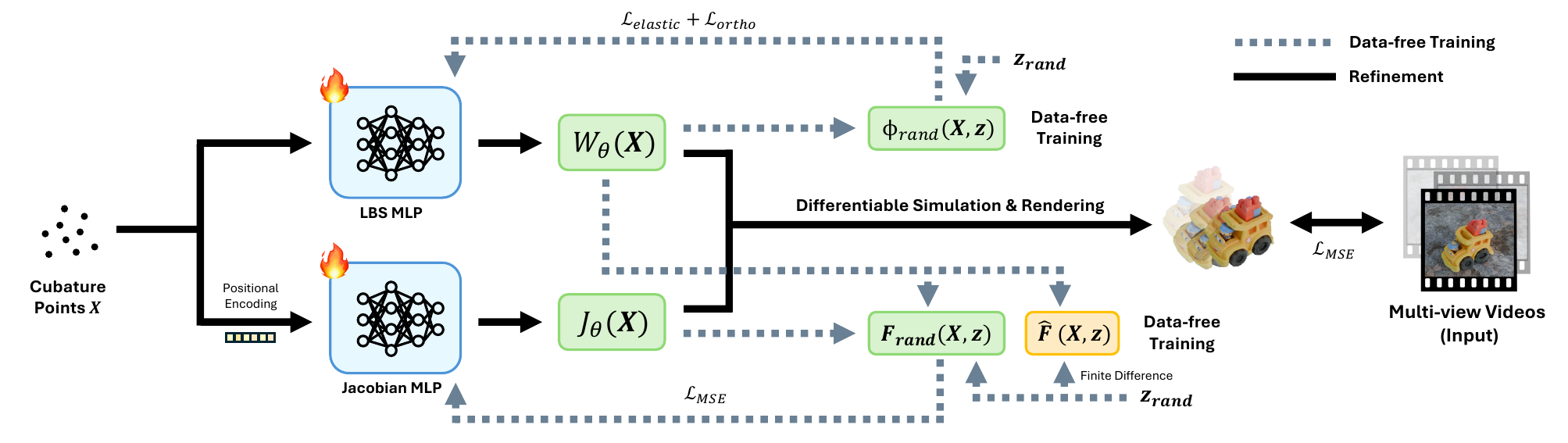}
  \caption{\label{fig:lbs} Network structure of LBS network and Jacobian network.}
\end{figure*}

\begin{table*}[!h]
\centering
\resizebox{\linewidth}{!}{
\begin{tabular}{c|c|cccccccccccc|c}
\toprule
\multicolumn{2}{c|}{} & backpack & bell & blocks & bus & cream & elephant & grandpa & leather & lion & mario & sofa & turtle & Mean \\ 
\midrule
\multirow{4}{*}{\rotatebox[origin=c]{90}{PSNR $\uparrow$}} & PAC-NeRF & 18.03 & 21.74 & 21.67 & 19.05 & 19.81 & 20.68 & 20.20 & 19.48 & 20.67 & 17.06 & 19.60 & 22.09 & 20.01 \\ 
& Spring-Gaus & 17.35 & 21.04 & 22.93 & 19.77 & 24.80 & 20.97 & 21.76 & 19.28 & 21.32 & 20.51 & 21.55 & 22.42 & 21.14 \\ 
& GIC & 18.22 & 19.33 & 18.26 & 20.23 & 23.94 & 21.50 & 20.45 & 20.76 & 18.53 & 21.17 & 23.38 & 22.60 & 20.70 \\ 
& Ours (full) & \textbf{26.59} & \textbf{27.26} & \textbf{31.29} & \textbf{25.64} & \textbf{33.85} & \textbf{27.95} & \textbf{24.09} & \textbf{31.11} & \textbf{25.89} & \textbf{26.54} & \textbf{27.82} & \textbf{30.81} & \textbf{28.24} \\
\midrule
\multirow{4}{*}{\rotatebox[origin=c]{90}{SSIM $\uparrow$}} & PAC-NeRF & 0.882 & 0.956 & 0.935 & 0.900 & 0.900 & 0.924 & 0.941 & 0.929 & 0.931 & 0.932 & 0.918 & 0.933 & 0.924 \\ 
& Spring-Gaus & 0.866 & 0.945 & 0.936 & 0.899 & 0.918 & 0.921 & 0.950 & 0.924 & 0.931 & 0.922 & 0.913 & 0.931 & 0.921 \\ 
& GIC & 0.879 & 0.938 & 0.918 & 0.903 & 0.901 & 0.924 & 0.943 & 0.937 & 0.917 & 0.930 & 0.920 & 0.934 & 0.920 \\ 
& Ours (full) & \textbf{0.940} & \textbf{0.964} & \textbf{0.971} & \textbf{0.935} & \textbf{0.951} & \textbf{0.955} & \textbf{0.953} & \textbf{0.975} & \textbf{0.949} & \textbf{0.951} & \textbf{0.942} & \textbf{0.973} & \textbf{0.955} \\
\midrule
\multirow{4}{*}{\rotatebox[origin=c]{90}{FoVVDP $\uparrow$}} & PAC-NeRF & 5.873 & 6.514 & 6.803 & 6.330 & 4.695 & 6.681 & 6.516 & 6.437 & 6.581 & 3.989 & 5.779 & 6.943 & 6.095 \\ 
& Spring-Gaus & 5.535 & 6.945 & 6.920 & 6.344 & 5.970 & 6.482 & \textbf{7.045} & 6.260 & 6.766 & 6.201 & 6.296 & 7.011 & 6.481 \\ 
& GIC & 5.959 & 6.129 & 5.947 & 6.394 & 5.805 & 6.921 & 6.930 & 6.696 & 6.023 & 7.029 & 6.905 & 6.939 & 6.473 \\ 
& Ours (full) & \textbf{8.331} & \textbf{7.640} & \textbf{8.921} & \textbf{7.898} & \textbf{8.900} & \textbf{8.398} & 7.020 & \textbf{9.053} & \textbf{7.659} & \textbf{8.015} & \textbf{7.912} & \textbf{9.000} &\textbf{8.229} \\
\bottomrule
\end{tabular}}
\caption{\label{tab:nvs} Quantitative comparison with previous methods on dynamic reconstruction (novel views).}
\end{table*}

\begin{table*}[!h]
\centering
\resizebox{\linewidth}{!}{
\begin{tabular}{c|c|cccccccccccc|c}
\toprule
\multicolumn{2}{c|}{} & backpack & bell & blocks & bus & cream & elephant & grandpa & leather & lion & mario & sofa & turtle & Mean \\ 
\midrule
\multirow{4}{*}{$\log(E)$} & PAC-NeRF & 3.28 & 1.08 & 4.02 & 3.30 & 3.22 & 3.05 & 2.99 & 1.20 & 2.34 & 3.37 & 0.20 & 1.94 & 2.50 \\ 
& GIC & 1.16 & 2.87 & 2.12 & 1.93 & 2.13 & 1.53 & \textbf{0.42} & 3.45 & 2.85 & 1.82 & 0.65 & 3.18 & 2.01 \\ 
& Ours (full) & \textbf{0.69} & \textbf{0.15} & \textbf{0.54} & \textbf{0.26} & \textbf{0.95} & \textbf{0.18} & 1.07 & \textbf{0.73} & \textbf{0.48} & \textbf{0.50} & \textbf{0.18} & \textbf{0.44} & \textbf{0.51} \\
\midrule
\multirow{4}{*}{$\nu$} & PAC-NeRF & 0.21 & 0.23 & 0.33 & 0.16 & 0.12 & 0.06 & 0.36 & 0.26 & 0.14 & 0.33 & 0.30 & \textbf{0.01} & 0.21 \\ 
& GIC & 0.11 & 0.24 & 0.29 & 0.18 & \textbf{0.08} & 0.24 & 0.26 & \textbf{0.02} & 0.14 & 0.22 & \textbf{0.01} & 0.08 & 0.16 \\ 
& Ours (full) & \textbf{0.10} & \textbf{0.10} & \textbf{0.07} & \textbf{0.06} & 0.11 & \textbf{0.05} & \textbf{0.02} & 0.06 & \textbf{0.05} & \textbf{0.06} & 0.08 & 0.02 & \textbf{0.06} \\
\bottomrule
\end{tabular}}
\caption{\label{tab:phys} Mean Absolute Error (MAE) among baselines and our method on physical property predictions.}
\end{table*}

\subsection{LBS and Jacobian Network}
\label{sec:lbs}
The implementation of the LBS and Jacobian network is visualized in
\cref{fig:lbs}. Specifically, the LBS network comprises 8 linear layers with a constant layer width of 64 and ELU~\cite{clevert2015fast} activation function. We observe that the neural Jacobian predominantly focuses on learning to predict high-frequency features, rather than the low-frequency signals typically modeled by the LBS prediction network. This insight motivates us to adopt a design for predicting the Jacobian that differs from the standard MLP architecture used in the LBS network, where we incorporate positional encoding into the input to capture the high-frequency features effectively. The input positions are embedded into a 512-dimensional space using positional encoding. The model comprises four residual blocks, each containing two linear layers. The first two residual blocks have a layer width of 512, while the last two have a layer width of 1024. The output is projected with a linear layer from the features. We use the GELU~\cite{hendrycks2016gaussian} activation function in the Jacobian network. We found that 4 blocks with positional encoding are sufficient to predict the Jacobian that is accurate enough for simulation, so we didn't scale up it further to save data-free training time. We report the speed-accuracy trade-off for different Neural Jacobian models in~\cref{tab:nj}. Note that the time cost of Neural Jacobian is only meaningful in its data-free training, which contains 10k iterations. It can be ignored in the joint optimization of Stage II.

The LBS and Jacobian networks are first trained in a data-free manner, supervised by randomly sampled $\mathbf{X}$ and $\mathbf{z}$, inspired by~\cite{modi2024simplicits, aigerman2022neural}. The LBS network is optimized by minimizing an elastic loss and orthogonal regularization loss. The Jacobian network is optimized by minimizing the L2 loss between the predicted deformation gradient $\mathbf{F(\mathbf{X}, \mathbf{z})}$ and the estimated $\mathbf{\hat F(\mathbf{X}, \mathbf{z})}$ from the finite difference.

The two networks are then jointly trained along with physical parameters according to the observed multi-view videos, where we only minimize the L2 loss between simulated animations and the observed multi-view videos, as described in Sec. 4.3 in the main paper.

\subsection{Boundary Condition Implementation}
We follow \cite{modi2024simplicits} to implement boundary conditions with incremental potential contact for handling collision, the constraints are formulated with barrier functions that provide extra potential energy. For example, our floor barrier in the dynamic reconstruction and the future state prediction task uses $E_f=10^5\times\Sigma_{i=1}^N [\max(0,h_f-h_i)]^2$ as potential energy, where $E_f$ is part of the external energy (See Eq. 2 in the main paper). Barrier functions can be very flexible in our method, and we provide more examples in~\cref{sec:bc}.
 
\section{More Results on Dynamic Reconstruction}
\label{sec:dy}

In \cref{sec:more}, we provide a comprehensive investigation by showcasing additional qualitative results of dynamic reconstruction and future state prediction across baselines, our Stage I model, and our full model. In \cref{sec:nvs}, we evaluate the performance of dynamic reconstruction on novel views. Additionally, we evaluate the prediction of physical properties $E$ and $\nu$ in \cref{sec:phys}.

\subsection{More Qualitative Results on Dynamic Reconstruction and Future States Prediction}
\label{sec:more}

\begin{figure*}[!h]
  \centering
  \includegraphics[width=\linewidth]{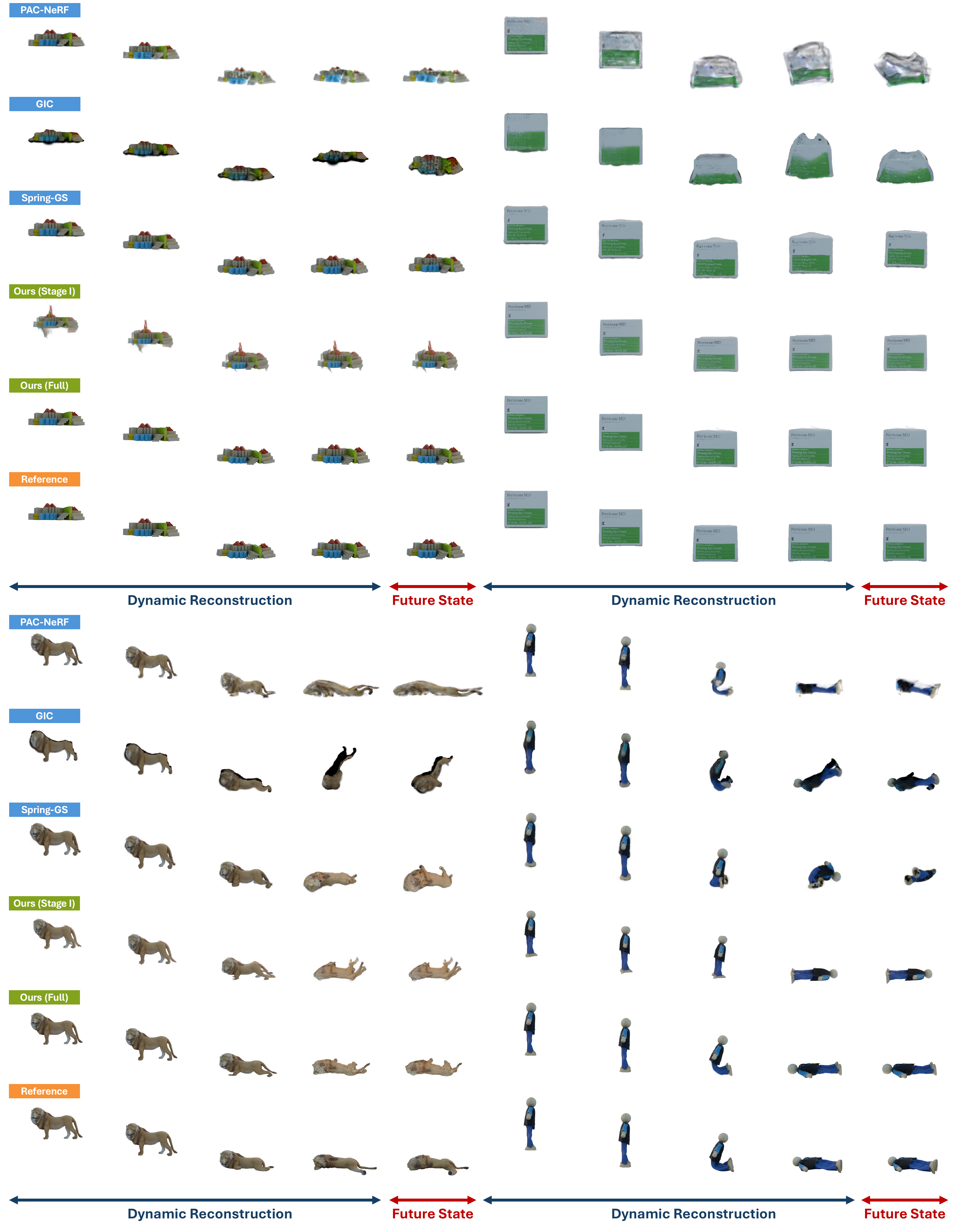}
  \caption{\label{fig:more1} More dynamic reconstruction results from the input videos.}
\end{figure*}

\begin{figure*}[!h]
  \centering
  \includegraphics[width=\linewidth]{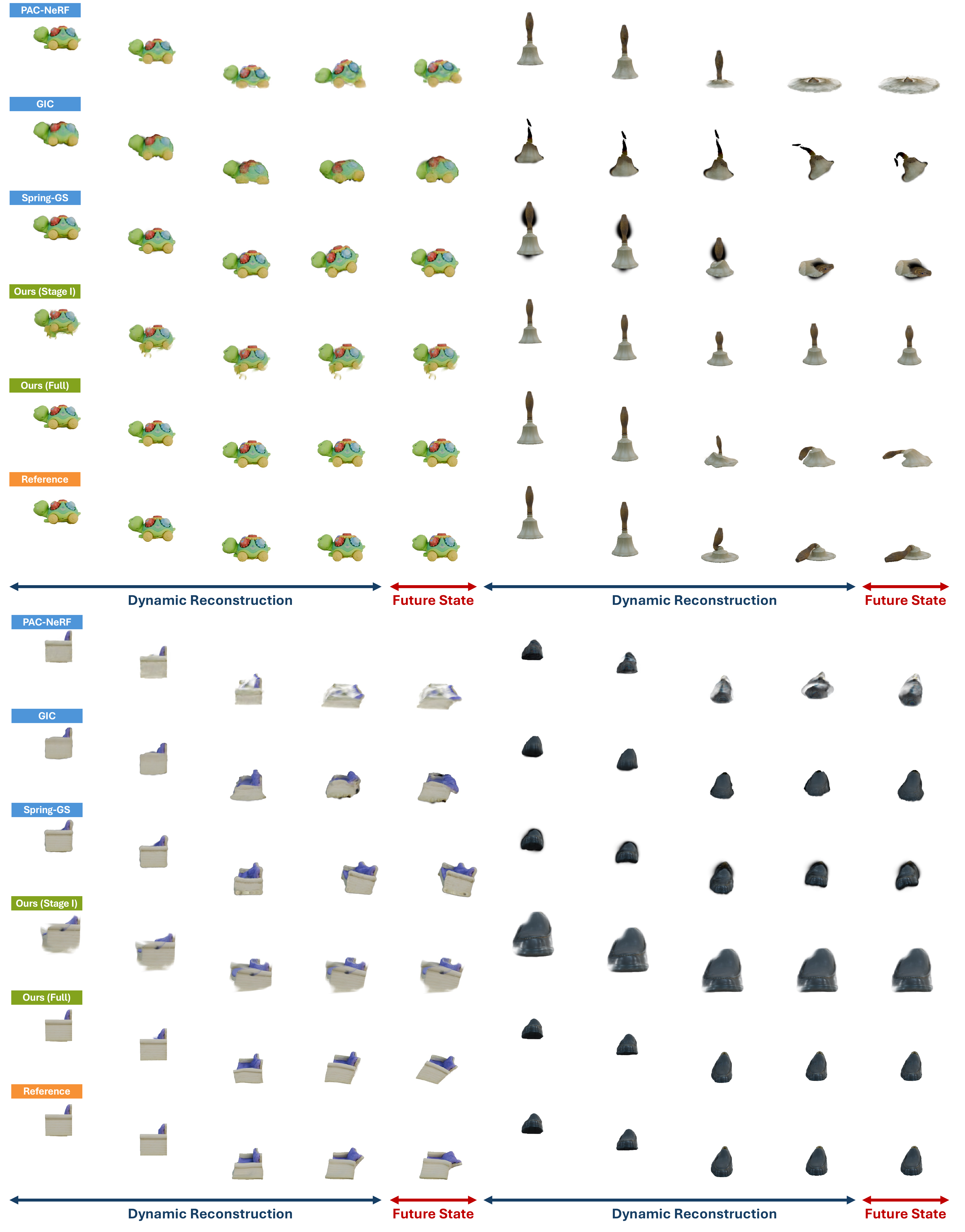}
  \caption{\label{fig:more2} More dynamic reconstruction results from the input videos.}
\end{figure*}

\begin{figure*}[htbp]
  \centering
  \includegraphics[width=\linewidth]{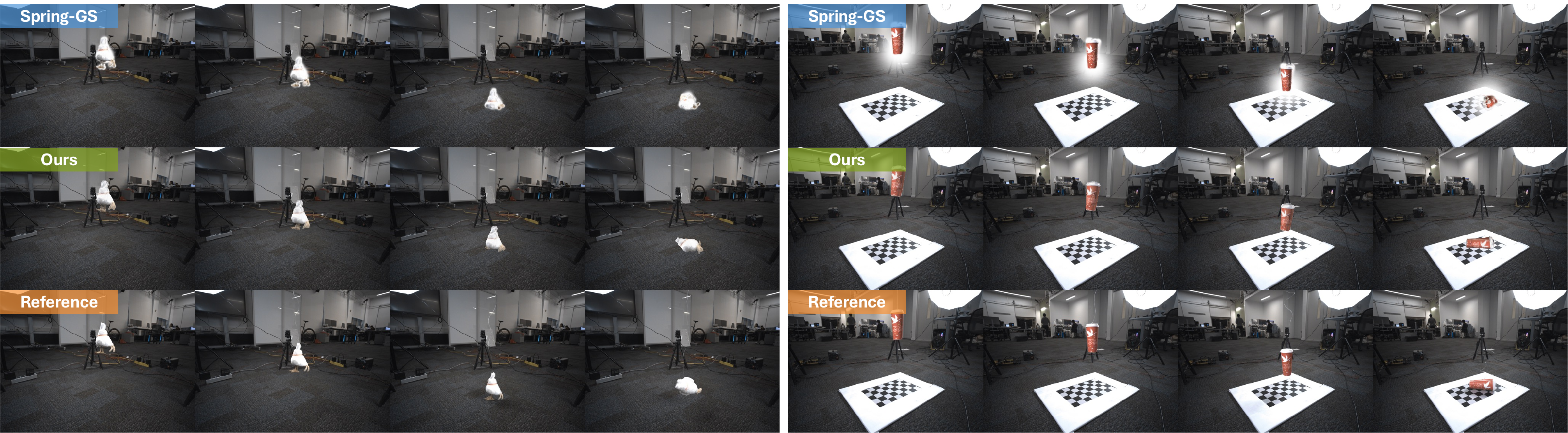}
  \caption{\label{fig:real2} More dynamic reconstruction results from real-world input videos.}
\end{figure*}

As illustrated in \cref{fig:more1} and \cref{fig:more2}, our model demonstrates remarkable physics-aware dynamic reconstruction quality compared to existing methods~\cite{li2023pac, cai2024gaussian, zhong2025reconstruction} that suffer from reconstructing blurry textures and incorrect dynamics due to the use of dynamic representations and symplectic solver. This is further evidenced by real-world test cases presented in \cref{fig:real2}, where the reconstruction results of the SOTA method Spring-Gaus~\cite{zhong2025reconstruction} collapses when hitting the ground plane,  while ours successfully capture the physical dynamics and produce higher realistic results.

\subsection{Evaluation on Novel View Synthesis}
\label{sec:nvs}
\begin{figure*}[!h]
  \centering
  \includegraphics[width=\linewidth]{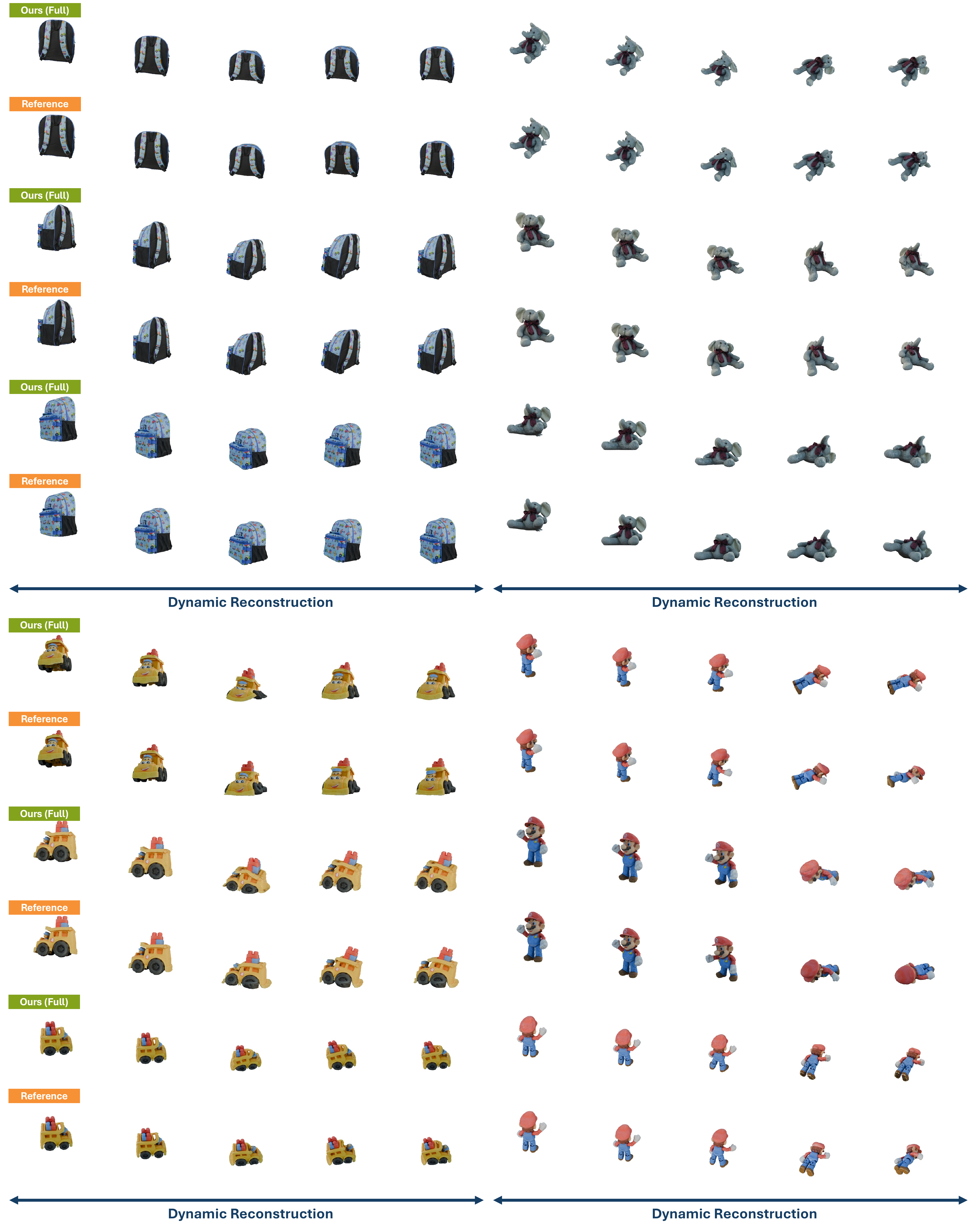}
  \caption{\label{fig:nvs} Novel view synthesis of the dynamic reconstruction results.}
\end{figure*}

We further evaluate the performance of our method on novel view synthesis by randomly sampling 6 novel views for each synthetic test case and evaluate the dynamic reconstruction performance among our method and baselines. We show qualitative results in \cref{fig:nvs} and quantitative results in \cref{tab:nvs}, where our method consistently outperforms other models.

\subsection{Evaluation on Physical Parameters Estimation}
\label{sec:phys}
Next, we evaluate the Mean Absolute Error (MAE) on the estimated $\log(E)$ and $\nu$ in the Neo-Hookean elastic model used by PAC-NeRF \cite{li2023pac}, GIC \cite{cai2024gaussian} and our method. 
As shown in \cref{tab:phys}, our method outperforms all the other approaches in most cases while showing its competitive performance on the remaining samples, which validates the effectiveness of our model on physical property estimation.

\section{Generalization Capability}
\label{sec:gen}

We provide more simulation results on changed materials in \cref{sec:mat} and provide additional simulation results on different boundary conditions in \cref{sec:bc}. 

\subsection{Generalized to Different Materials}
\label{sec:mat}
\begin{figure*}[!h]
\vspace{-6mm}
  \centering
  \includegraphics[width=\linewidth]{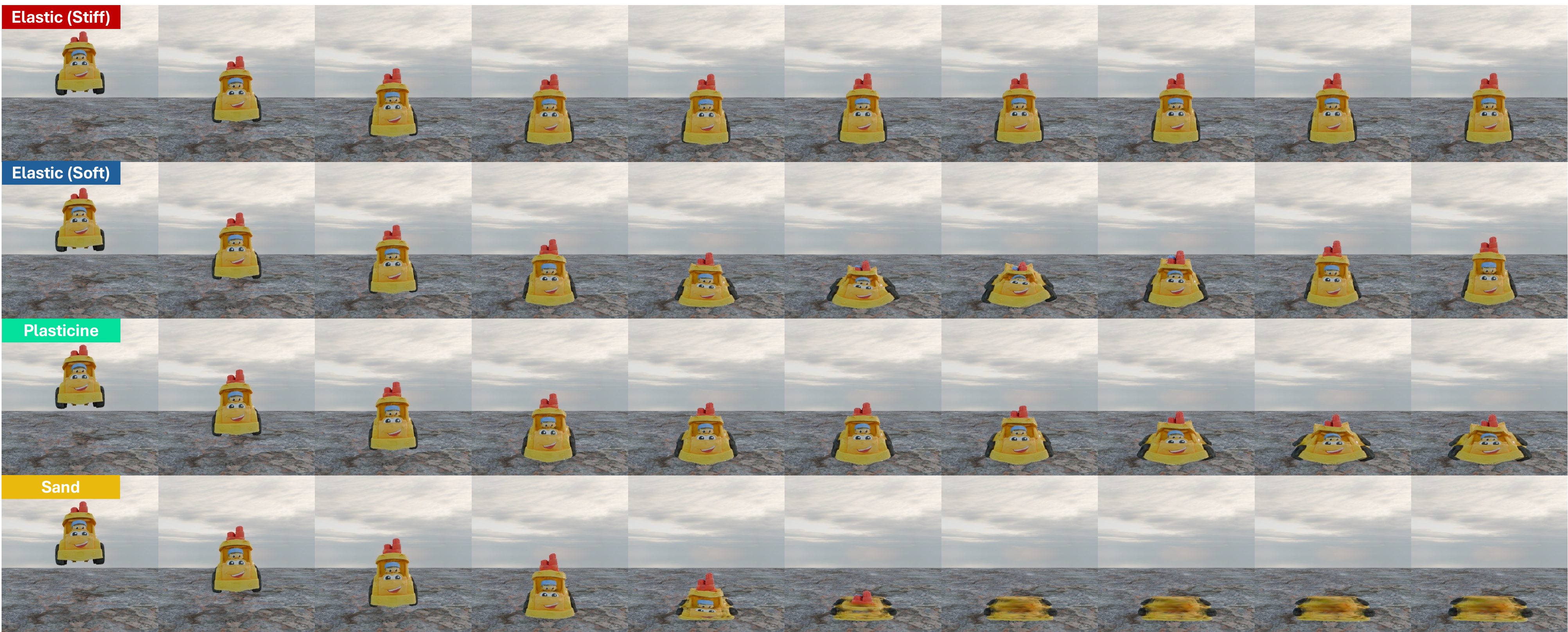}
  \caption{Simulation with different materials. We use $E=10^7, \nu=0.49$ for the stiff elastic and $E=8000, \nu=0.4$ for the soft elastic. In Plasticine material $\tau_{Y}$ is set to $500$ and in Sand material $\theta_f$ is set to $10 \degree$.}
  \label{fig:mat}
  \vspace{-1mm}
\end{figure*}

Although our method mainly focuses on reconstructing elastic objects in this paper, our framework can be generalized to materials characterized by various constitutive models. Here, we show simulation results regarding three different materials: Elasticity, Plasticine, and Sand following \cite{cai2024gaussian, li2023pac}. The qualitative results are shown in \cref{fig:mat}, where different materials are simulated precisely as our method is combined with different constitutive models effectively.

\begin{figure*}[!h]
  \centering
  \begin{subfigure}[b]{\linewidth}
    \centering
    \includegraphics[width=\linewidth]{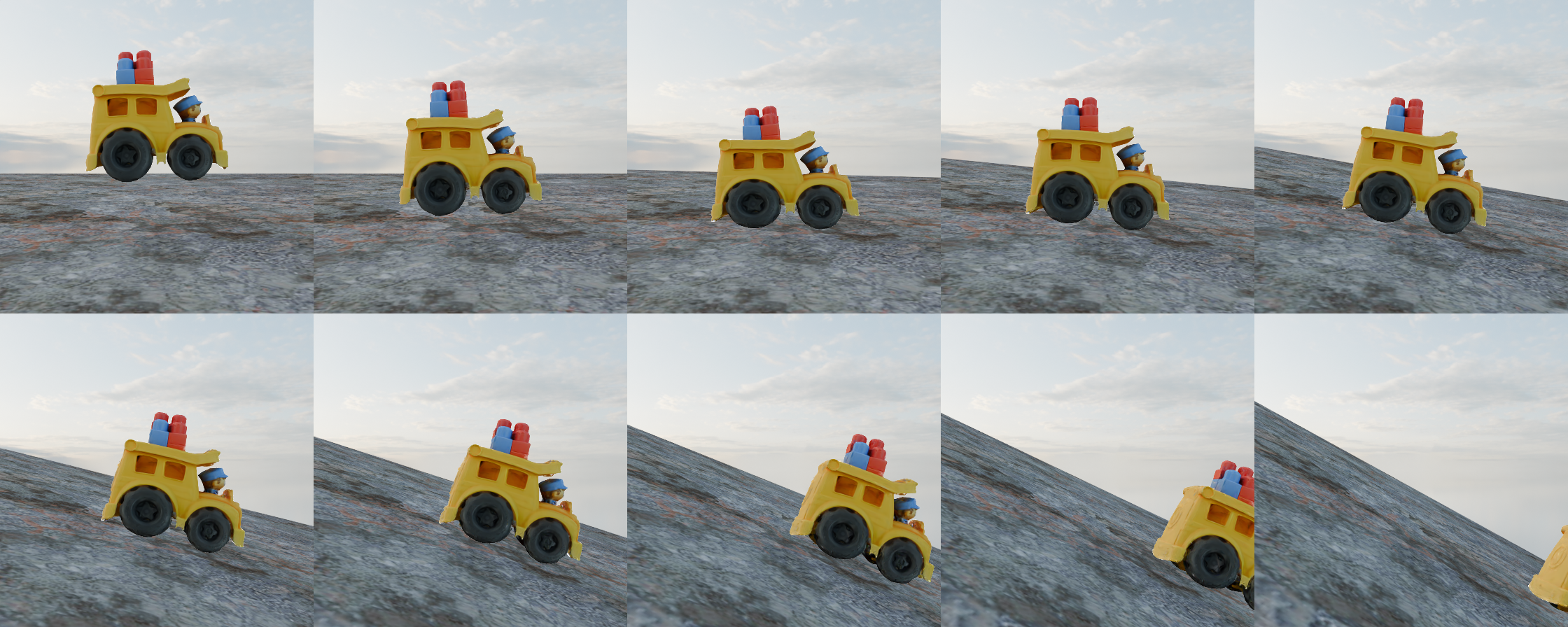}
    \caption{\textit{A bus slides at a \textbf{moving} floor.}}
    \label{fig:bc_a}
  \end{subfigure}
  \hfill
  \begin{subfigure}[b]{\linewidth}
    \centering
    \includegraphics[width=\linewidth]{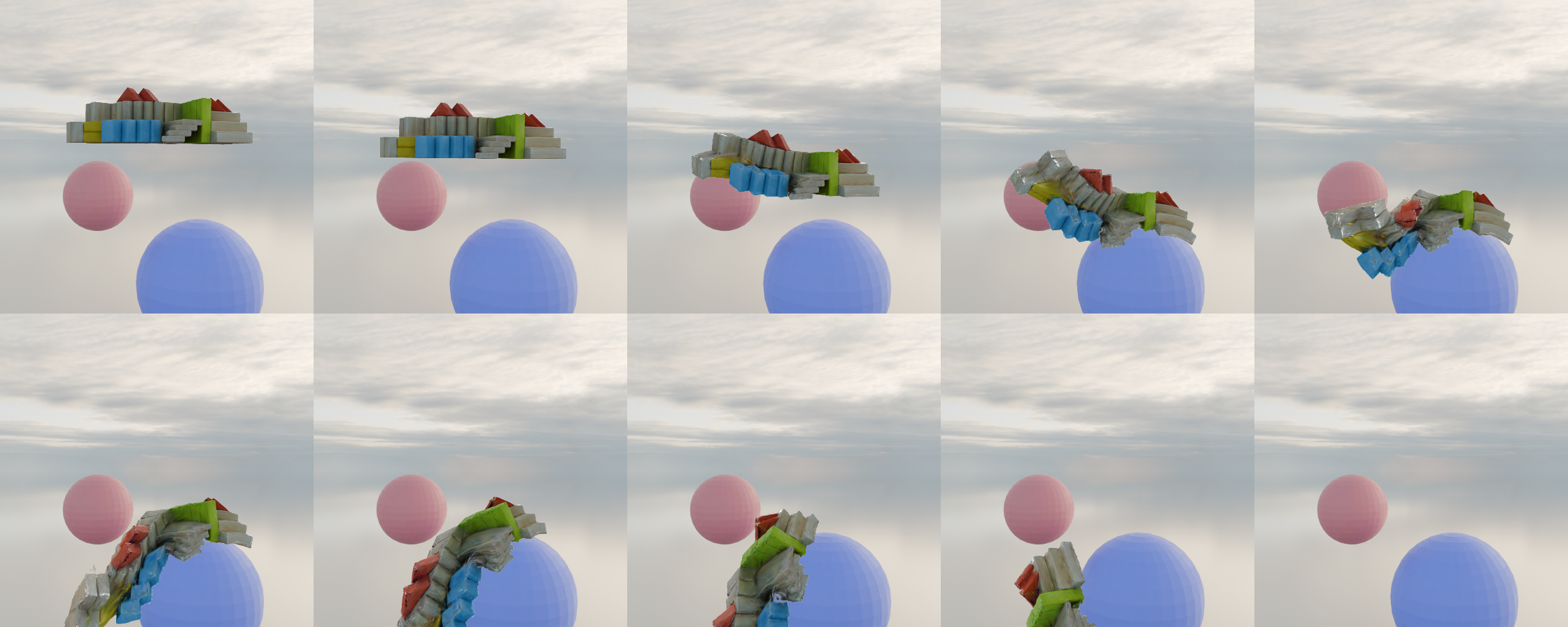}
    \caption{\textit{Blocks drop on the \textbf{balls}.}}
    \label{fig:bc_b}
  \end{subfigure}
  \vspace{-7mm}
  \caption{\label{fig:bc} Simulation results based on different boundary conditions.}
  \vspace{-10mm}
\end{figure*}

In order to compute the potential energy $E_{potential}$ for simulation, we derive the corresponding energy density function $\Psi(\mathbf{F})$ for each constitutive model below.

\paragraph{Elasticity.} The energy density function can be formulated as
\begin{equation}
    \Psi(\mathbf{F})=\frac{\mu}{2} [tr(\mathbf{F}^{\top}\mathbf{F})-d]-\mu \ln(J)+\frac{\lambda}{2} \ln^2(J)
\end{equation}
where $d=3$ is the space dimension, $\mathbf{F}$ is the deformation gradient and $J$ is the determinant of $\mathbf{F}$, $\mu$ and $\lambda$ are Lam\'e parameters related to Young’s modulus $E$ and Poisson’s ratio $\nu$:
\begin{equation}
    \mu=\frac{E}{2(1+\nu)} \quad \lambda=\frac{E \nu}{(1+\nu)(1-2\nu)}
\end{equation}

\paragraph{Plasticine}
Plasticine material is modeled with a combination of Saint Venant-Kirchhoff Model (StVK) and von Mises return mapping function. The energy density function of StVK can be formulated as
\begin{equation}
    \Psi(\mathbf{F})=\mu [tr(\mathbf{G}^2)] + \frac{\lambda}{2} [tr^2(\mathbf{G})]
\end{equation}
where $\mathbf{G}=\frac{1}{2}(\mathbf{F}^{\top}\mathbf{F}-d)$ is the Green strain. The von-Mises return mapping function projects the deformation gradient back onto the boundary of the elastic region according to the von-Mises yielding condition. The mapping function can be formulated as
\begin{equation}
    \mathcal{Z}(\mathbf{F})=
    \left\{ \begin{array}{ll}
     \mathbf{F} & \delta \gamma \le 0 \\ 
     \mathbf{U} \exp(\mathbf{\epsilon}-\delta \gamma \frac{ \mathbf{\hat \epsilon}}{\|\mathbf{\hat \epsilon}\|}) \mathbf{V}^{\top} & \rm{otherwise} \\
    \end{array}\right.
\end{equation}
where $\mathbf{F}=\mathbf{U}\mathbf{\Sigma}\mathbf{V}^{\top}$ is the singular value decomposition (SVD) of $\mathbf{F}$, $\mathbf{\epsilon}=\log(\mathbf{\Sigma})$ is the Hencky strain, $\mathbf{\hat \epsilon}=\mathbf{\epsilon}-\mathbf{\bar \epsilon}$ is the normalized Hencky strain and $\delta \gamma=\|\mathbf{\hat \epsilon}\|-\frac{\tau_{Y}}{2\mu}$ is von-Mises yielding condition with the yield stress $\tau_{Y}$ as a physical parameter.

\paragraph{Sand}
Similar to the Plasticine material, we also use StVK as the constitutive model and apply its energy density function to the Sand material. The difference is that we use Drucker-Prager yield criteria instead of von-Mises yield criteria. The mapping function can be formulated as
\begin{equation}
    \mathcal{Z}(\mathbf{F})=
    \left\{ \begin{array}{ll}
     \mathbf{U}\mathbf{V}^{\top} & tr(\mathbf{\epsilon}) > 0 \\ 
     \mathbf{F} & \delta \gamma \le 0,~tr(\mathbf{\epsilon}) \le 0 \\ 
     \mathbf{U} \exp(\mathbf{\epsilon}-\delta \gamma \frac{ \mathbf{\hat \epsilon}}{\|\mathbf{\hat \epsilon}\|}) \mathbf{V}^{\top} & \rm{otherwise} \\
    \end{array}\right.
\end{equation}
where $\delta \gamma=\|\mathbf{\hat \epsilon}\|_F+\alpha\frac{(d \lambda + 2\mu)tr(\mathbf{\epsilon)}}{2\mu}$ is the yield stress, $\alpha=\sqrt{\frac{2}{3}}\frac{2 \sin \theta_{f}}{3-\sin \theta_{f}}$ and $\theta_{f}$ is the friction angle. 

\subsection{Generalized to Complex Boundary Conditions}
\label{sec:bc}
In this section, we demonstrate that the reconstruction results of our method, \name, integrate seamlessly into the simulation of various animations under complex boundary conditions. Two examples are presented in \cref{fig:bc}, highlighting \name's ability to generate high-quality animations across diverse boundary scenarios.

\end{document}